\documentclass[twoside]{article}

\usepackage{lipsum} 

\usepackage[round]{natbib}
\usepackage[sc]{mathpazo} 
\usepackage[T1]{fontenc} 
\usepackage{microtype} 

\usepackage[hmarginratio=1:1,top=32mm,columnsep=20pt]{geometry} 
\usepackage{multicol} 
\usepackage[hang, small,labelfont=bf,up,textfont=it,up]{caption} 
\usepackage{booktabs} 
\usepackage{float} 
\usepackage{lettrine} 
\usepackage{paralist} 

\usepackage{abstract} 

\usepackage{titlesec} 
\renewcommand\thesection{\Roman{section}} 
\renewcommand\thesubsection{\Roman{subsection}} 
\titleformat{\section}[block]{\large\scshape\centering}{\thesection.}{1em}{} 
\titleformat{\subsection}[block]{\large}{\thesubsection.}{1em}{} 

\usepackage{fancyhdr} 
\usepackage{graphicx}
\DeclareGraphicsExtensions{.jpg}
\usepackage{caption}
\usepackage{subcaption}
\usepackage{amsmath}
\usepackage[hidelinks]{hyperref}

\title{\vspace{-15mm}\fontsize{25pt}{10pt}\selectfont\textbf{ Characterization of Large-Scale Functional Brain Networks During  Ketamine-Medetomidine Anesthetic Induction   }} 

\author{
\large
\textsc{Eduardo C. Padovani}
\thanks{Email: \texttt{eduardo.padovani@alumni.usp.br}}\\[2mm] 
\normalsize Universidade de S\~ao Paulo \\ 
\normalsize 
\vspace{-5mm}
}
\date{}

\begin{document}

\maketitle 

\thispagestyle{fancy} 


\begin{abstract}
   
\noindent \ Several experiments provide evidence that specialized brain regions functionally interact and reveal that the brain processes and integrates information in a specific and structured manner. Networks can be applied to model brain functional activities, providing means to characterize and quantify this structured form of organization. Reports substantiate that different physiological states or diseases that affect the central nervous system may be associated with alterations in these networks, which might be reflected in graphs of different architectures. However, the relationship between their structure and the organism's distinct physiological conditions is poorly comprehended. Therefore, experiments that estimate the functional neural networks of subjects exposed to different controlled conditions are highly relevant. Within this context, this research has sought to model large-scale functional brain networks during an anesthetic induction process. The experiment was based on intra-cranial recordings of the neural activities of an old-world macaque of the species Macaca fuscata. Neural activity was recorded during a Ketamine-Medetomidine anesthetic induction process, and networks were estimated sequentially in five-second intervals. One and a half minutes after administering the anesthetics, changes occurred in various network properties, revealing a transition in the network architecture. During general anesthesia,  functional connectivity and network integration capabilities were reduced at both local and global levels. Additionally, it has been verified that the brain shifted to a highly specific and dynamic state. The results provide empirical evidence and report the relationship between the induced state of anesthesia and functional network properties, contributing to the elucidation of novel aspects of the neural correlates of consciousness.

\end{abstract}


\begin{multicols}{2} 

\section{Introduction}
\lettrine[nindent=0em,lines=2]{O}ne of the main goals of neuroscience is to comprehend how the brain works by understanding how cognitive abilities or physiological states of the organisms are related to neural processes involving functional interactions of various brain structures. Although neuroscience is a well-developed and consolidated science, understanding has not yet been reached at this level, and possibly the scientific community will still need several years to comprehend the brain to this extent. 
In recent times, due to the influence of other fields of science, such as the physics of complex systems, allied with the constant realization that neural activities involve the simultaneous participation of many distinct cortical areas, a new perspective has been gaining strength among the community of neuroscientists. This perspective aims to understand the brain as a whole from a system point of view and considers as fundamental the comprehension of the way functional interactions occur among distinct brain structures to understand how the brain is able to perform its activities \citep{bullmore2009complex}.\
Based on this perspective, researchers have been using concepts and tools from the \textit{Modern Network Science} \citep{strogatz2001exploring,newman2003structure} to model, characterize, and study the brain.

The \textit{Modern Network Science} is a highly interdisciplinary
 field of science aimed at understanding the functioning,
 behavior, and evolution of complex systems, based on properties
 of their structure, that is, the specific way in which the elements of the system establish interactions \citep{mitchell2009complexity}.
One of the main uses of complex networks in neuroscience is the modeling of functional interactions established among distinct cerebral regions by using the mathematical structure of graphs. Assembling those networks in which nodes represent cortical areas and the edges functional interactions \citep{friston1993functional,friston1994functional} established between these regions, several procedures and analyses can be applied to evaluate and study the properties of these networks. By using complex networks measures \citep{rubinov2010complex}, it is possible to estimate how the connectivity is shaped, the organization at local and global levels, the nodes that perform great influence in the integration of several regions, and the mediation of the information flow \citep{bullmore2009complex,sporns2011networks,
stam2012organization}. It is believed that these networks reflect the specific ways in which the processing and integration of information are performed among distinct brain structures. There is also a consensus that different cognitive demands or disorders that affect the central nervous system may be associated with specific network configurations  \citep{stam2007graph,sporns2011networks}.

 \vspace{-0.7\baselineskip}

\subsubsection*{General Anesthesia}
 
 \vspace{-0.5\baselineskip}
 
Understanding general anesthesia is highly relevant to both medicine and neuroscience. Given its importance and the fact that general anesthesia's underlying mechanisms are not well understood, the \textit{Science Magazine} has pointed out the elucidation of processes and mechanisms involving general anesthesia as one of the 125 most important open questions in science \citep{kennedy2005don}.

Anesthetic agents are small molecules that can interact with and modulate the activity of specific ionic channel proteins in neurons, promoting dramatic physiological alterations in the organism.
General anesthesia is a drug-induced physiologically stable and reversible state characterized by analgesia, amnesia, immobility, and loss of consciousness \citep{schwartz2010general}. The anesthetic drug's pharmacological effects are reasonably known and described. However, the neurophysiological mechanisms that underlie sedation and loss of consciousness are not yet well understood \citep{schwartz2010general,lewis2012rapid}. Besides their undeniable importance and usefulness in clinical medicine, these drugs can also constitute tools of great value to neuroscience. As anesthetic agents are able to induce different levels of consciousness in a stable and reproductive manner \citep{uhrig2014cerebral}. They can be used as experimental tools, providing means for the study of consciousness and the neural correlates of consciousness, thus offering possibilities to investigate fundamental processes and phenomena that occur in the brain \citep{hameroff1998toward}.

There are several theories about anesthesia \citep{flohr1995information,alkire2000toward,
mashour2004consciousness,john2005anesthetic}, as well as theories of consciousness based on the interface between consciousness and anesthesia \citep{hameroff2006entwined,mashour2006integrating}. There are also several hypotheses and reports on how anesthetic agents lead to the loss of consciousness, based on the depression of cerebral functions \citep{alkire1995cerebral,alkire1998quantitative}, on the reduction of functional interactions between brain areas \citep{white2003impaired,imas2005volatile}, on the fragmentation of neural networks \citep{lewis2012rapid}, and others.\\

\vspace{-0.5\baselineskip}

Within this context of several theories, hypotheses, and reports, the present study sought to investigate the alterations in the organization of functional brain activities that occur at the onset of the anesthetic induction by using concepts and tools of the \textit{Modern Network Science}. In order to perform those analyses, macaque's functional brain networks were estimated serially over time during an anesthetic induction process. Complex network measures were employed to characterize and compare the graphs estimated at different instants of time throughout the experiment.


\section{Methods}

  \subsubsection*{Database} 
  
In order to study the effects induced by general anesthesia in the functional brain networks, we have analyzed an ECoG electrophysiological records database provided by the Adaptive Intelligence laboratory at the \texttt{RIKEN Brain Science Institute}, Saitama, Japan. The database was respective to an experiment involving a  \textit{Ketamine-Medetomidine} anesthetic induction in a non-human primate animal model subject of the species \textit{Macaca fuscata}, who had an MDR-ECoG matrix chronically implanted in the subdural space extensively covering the left brain hemisphere's lateral cortical surface and also the frontal and occipital medial walls \citep{nagasaka2011multidimensional}. The database is available in the public domain at (\texttt{http://neurotycho.org}). For further information see \citep{nagasaka2011multidimensional}.

\vspace{-0.5\baselineskip}

\subsubsection*{Recording Technique}
 
The Multidimensional Recording Electrocorticogram (MDR-ECoG) is considered the most advanced and balanced technology to record cortical electrophysiological activity \citep{yanagawa2013large}. It can sample neural signals at temporal resolutions higher than 1KHz, and a spatial resolution of up to 3mm, offering concomitantly high spatial and temporal resolution. Furthermore, once the electrodes are positioned right over the cortex, beneath the dura mater, the MDR-ECoG is also regarded to provide reliable records of cortical electrophysiological activity along with low levels of noise \citep{yanagawa2013large,fukushima2014electrocorticographic}.

 \vspace{-0.5\baselineskip}
 
\subsubsection*{Animal Model Subject}
 
Experimentally, an old-world monkey of the species \textit{Macaca fuscata} was used as an animal model. This species of macaque has considerable anatomical and evolutionary similarities with humans, making them an excellent platform for the study of the human brain \citep{iriki2008neuroscience}.

\subsubsection*{Anesthetic Agents}

The neural records database is respective to an experiment that involved a \textit{Ketamine} and \textit{Medetomidine} anesthetic induction in a macaque subject.
 
Ketamine is a drug that induces an anesthetic state characterized by the dissociation between the thalamocortical and limbic systems \citep{bergman1999ketamine}. It acts as a non-competitive antagonist of the receptor N-methyl-D-aspartate \citep{green2011clinical}. Medetomidine (an agonist of the alpha-2 adrenergic receptor) was combined with Ketamine to promote muscular relaxation \citep{young1999short}. The antagonist of  Medetomidine, Atipamezole, was used to trigger and promote the recovery process \citep{young1999short}.

\subsubsection*{Neural Connectivity Estimator}
 
As a neural connectivity estimator, a methodology based on Granger causality in the frequency domain \citep{granger1969investigating,seth2007distinguishing} was used to infer the statistical dependencies between the time series of the electrodes. When applied in neuroscience, Granger causality provides an estimate of the information flow from one cortical area to another \citep{seth2007distinguishing,seth2010matlab}.

\subsubsection*{Experimental Procedures - Summary of Steps}

The modeling of functional neural activities performed in this study followed these steps:

\begin{enumerate}

\item  The database is respective to neural registers recorded by the MDR-ECoG technique. The matrix of ECoG electrodes continuously covered the entire left brain hemisphere and parts of the cortical medial walls.
 
\item Each electrode of the ECoG array was considered a vertex of the network and represented the respective cortical area in which it was positioned.
 
\item  A neural connectivity estimator of Granger causality in the frequency domain was used to estimate the association values between the electrodes registers (time series).
 
\item An adjacency matrix was assembled, containing all the pairwise association values between the nodes.
 
\item The characterization of the topology of the estimated networks was performed using complex network measures.

\end{enumerate}
 
Networks were estimated sequentially at five-second intervals in five physiological frequency bands throughout the experiment. In each frequency band, all the networks in the sequence were obtained using the same procedures and parameters. Thus, the alterations observed in the properties of distinct networks over the course of the experiment came only from the differences in the records of neural activity.\\

\vspace{-1.5\baselineskip}

\subsection{Database - Anesthetic Induction Experimental Procedures}{

The database was recorded during an experiment conducted according to the following experimental procedures:\\

 The monkey was seated in a proper chair with its head and arms restrained. The neural activity started to be recorded while the monkey was awake and with open eyes. After that, the eyes were covered with a patch to prevent visual evoked responses. After about 10 minutes, a Ketamine-Medetomidine cocktail (5.6mg/Kg of Ketamine + 0.01mg/Kg of Medetomidine) was injected intramuscularly to induce anesthesia. The loss of consciousness point (LOC) was set at the time when the monkey no longer responded to external stimuli (touching the nostrils or opening the hands). After establishing the LOC, neural activity was recorded for about 25-30 minutes. Heart rate and breathing were monitored throughout the entire experiment. For further information, see (\texttt{http://neurotycho.org}).
 
}

\subsection{Signal Processing and Granger Causality in the Frequency Domain}

\vspace{0.5\baselineskip}

\subsubsection*{Data Processing}

\vspace{0.5\baselineskip}

\begin{enumerate}
 
\item A reject-band \texttt{IIR-notch} filter was used to attenuate components of the signal at 50Hz.
 
\item The signal was down-sampled from 1KHz to 200Hz.
 
\item The signal was divided into windows of 1000 points (equivalent to a five-second recording of neural activity).
 
\item For each of the 128-time series, the trend was removed, and the average was subtracted.
 
\item To verify the stationary condition of the time series, the \textit{KPSS} \citep{kwiatkowski1992testing} and the \textit{ADF} [Augmented Dickey Fuller] \citep{hamilton1989new} tests were applied.
  
\end{enumerate}

\subsubsection*{Libraries Used}

For the computation of association values using Granger causality in the frequency domain, were used with some adaptations the following libraries: \texttt{MVGC GRANGER TOOLBOX}, developed by Ph.D. Anil Seth (Sussex University, UK), described in \citep{seth2010matlab}, available at \texttt{www.anilseth.com}, and the \texttt{BSMART toolbox} (\textit{\textbf{B}rain-\textbf{S}ystem for \textbf{M}ultivariate \textbf{A}uto\textbf{R}egressive \textbf{T}imeseries} \texttt{toolbox}) described in \citep{cui2008bsmart} and available at \texttt{www.brain-smart.org}.

\newpage

\subsubsection*{Computation of Causal Interactions}

\begin{enumerate}
 
\item Model Order:
 
To find the model order (number of observations to be used in the regression model), the model's selection criteria from Akaike (AIC) and Bayes/Schwartz (BIC) were used.
Both methods returned the order of the model equal to seven.
 
\item Causal Interactions
 
At each window of 1000 points, Granger causality in the frequency domain interactions were pair-wise computed among the 128-time series by the use of the function \texttt{cca\_pwcausal()} (\texttt{MVGC GRANGER TOOLBOX}).

\item Frequency Bands
 
 Granger causality interactions were calculated in five physiological frequency bands: Delta (0-4Hz), Theta (4-8Hz), Alpha (8-12Hz), Beta (13-30Hz), and Gamma (25-100Hz). 
 
The interaction values obtained were saved in adjacency matrices.

\end{enumerate}

\subsubsection*{Graphs and Networks}

\begin{enumerate}

\item \textbf{Assemble Networks}

For each sequence of graphs respective to a frequency band, a threshold was chosen, and only the interactions with magnitude values higher than this threshold were considered edges of the graphs.

\begin{itemize}
\item Delta (0-4Hz), threshold = $0.8$
\item Theta (4-8Hz), threshold = $0.5$
\item Alpha (8-12Hz), threshold = $0.5$
\item Beta (13-30Hz), threshold = $1.0$
\item Gamma (25-100Hz), threshold = $2.5$
 
\end{itemize}

As discussed in \citep{bullmore2009complex,sporns2011networks}, scientists can rely on various criteria to determine this parameter. In the present study, due to experimental conditions, each sequence of networks contained graphs with distinct connectivity. Thresholds were selected in such a way as to prevent graphs with lower connectivity in each sequence from presenting many disconnected parts or vertices, which could potentially introduce distortions in the analysis.\\

After obtaining non-weighted graphs, the directions of the edges were removed, resulting in undirected and non-weighted networks.

 \item \textbf{Analysis of Topology}
 
Network measures \citep{rubinov2010complex} were used to characterize the graph's topology.

\end{enumerate}


\section{Results}

Alterations in the topology of the functional brain networks were observed along the anesthetic induction process, verifying the occurrence of changes over the distinct measures used to characterize the networks. Those results reveal differences in how cortical areas functionally interact regarding awake conditions and the induced state of anesthesia.

About one and a half minutes after administering the anesthetic cocktail, abrupt changes were observed over several network properties, revealing the existence of a rapid transition between the two states (awake and anesthetized). 
Alterations were also observed after the macaque was blindfolded, demonstrating that different stimuli presented to the animal were able to promote changes in its functional neural network structure.

\vspace{-0.8\baselineskip}

\subsection{Average Degree}

Significant alterations in the network's average degree were observed on the five frequency bands analyzed during the experiment (see \hyperlink{FIGURE1}{$Figure \cdot 1$}; \hyperlink{TABLE1}{$Table \cdot 1$}).

Regarding the transition between resting state with eyes open and blindfolded conditions, changes in the network's average connectivity were noted in some frequency bands (see \hyperlink{FIGURE1}{$Figure \cdot 1$}). After placing a patch over the eyes, the average degree presented a tendency to increase and also to display a higher variation.

Noticeable changes in the average degree of the graphs were also observed in the transitional period between the awake and anesthetized states (see \hyperlink{FIGURE1}{$Figure \cdot 1$}); the average degree presented a considerable reduction and was less prone to variation. Under general anesthesia conditions, the graphs assumed a tendency to possess the same connectivity over time.


\end{multicols}

\begin{figure}[!h]
\begin{subfigure}{.5\textwidth}
  \centering
  \includegraphics[width=1\linewidth]{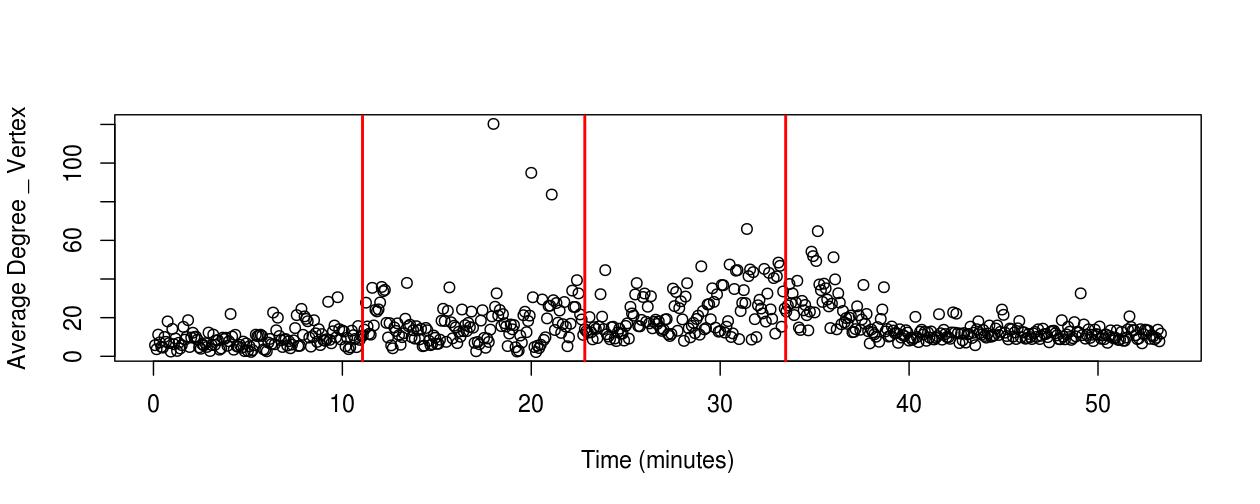}
  \caption{Delta 0-4Hz}
  \label{fig:sfig1}
\end{subfigure}%
\begin{subfigure}{.5\textwidth}
  \centering
  \includegraphics[width=1\linewidth]{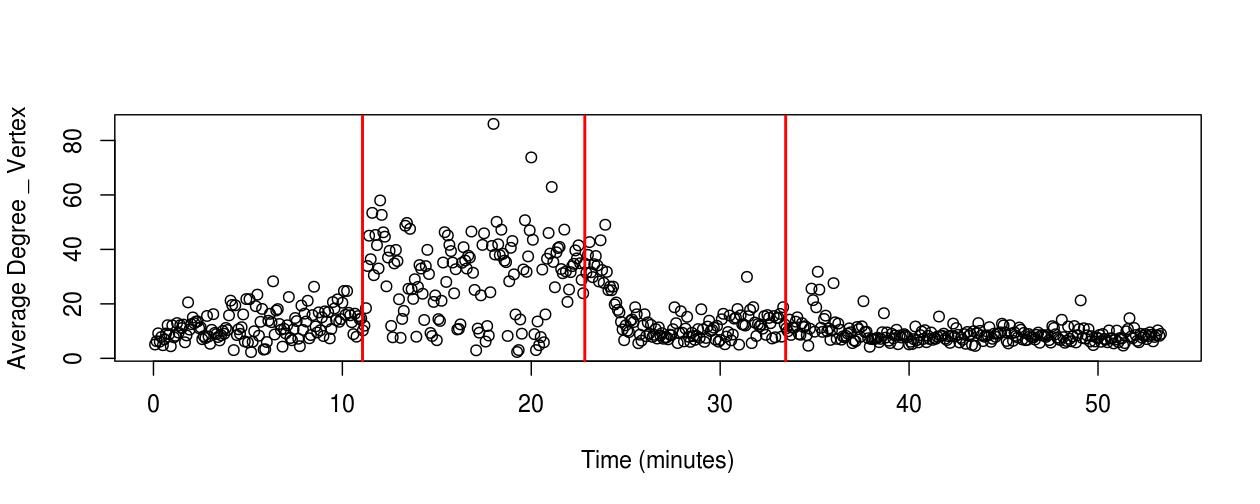}
 \caption{Theta 4-8Hz}
  \label{fig:sfig2}
\end{subfigure}\\
\centering
\begin{subfigure}{.5\textwidth}
\includegraphics[width=1\linewidth]{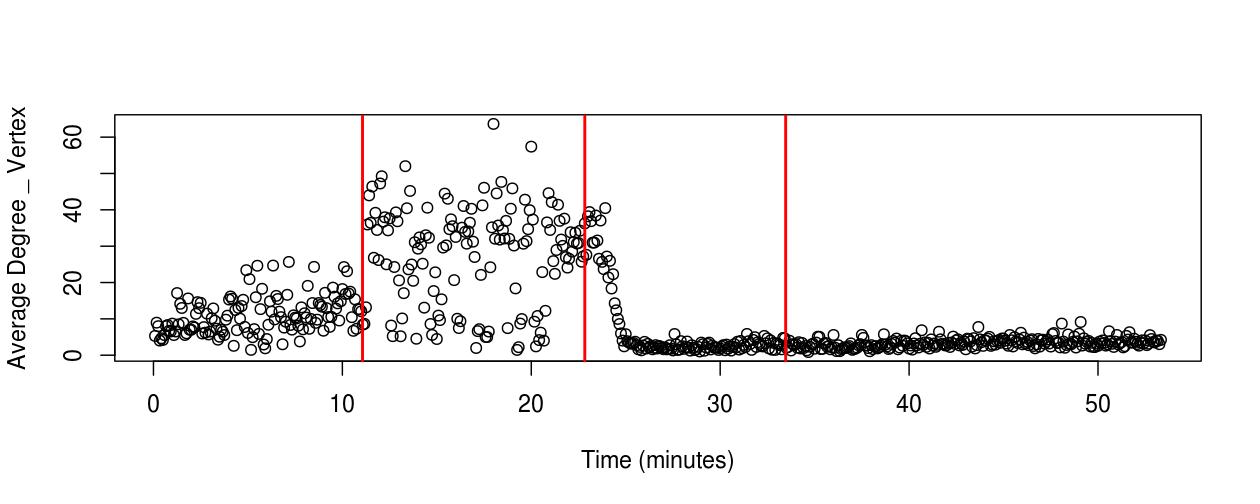}
  \caption{Alpha 8-12Hz}
  \label{fig:sfig3}
\end{subfigure}%
\begin{subfigure}{.5\textwidth}
  \centering
  \includegraphics[width=1\linewidth]{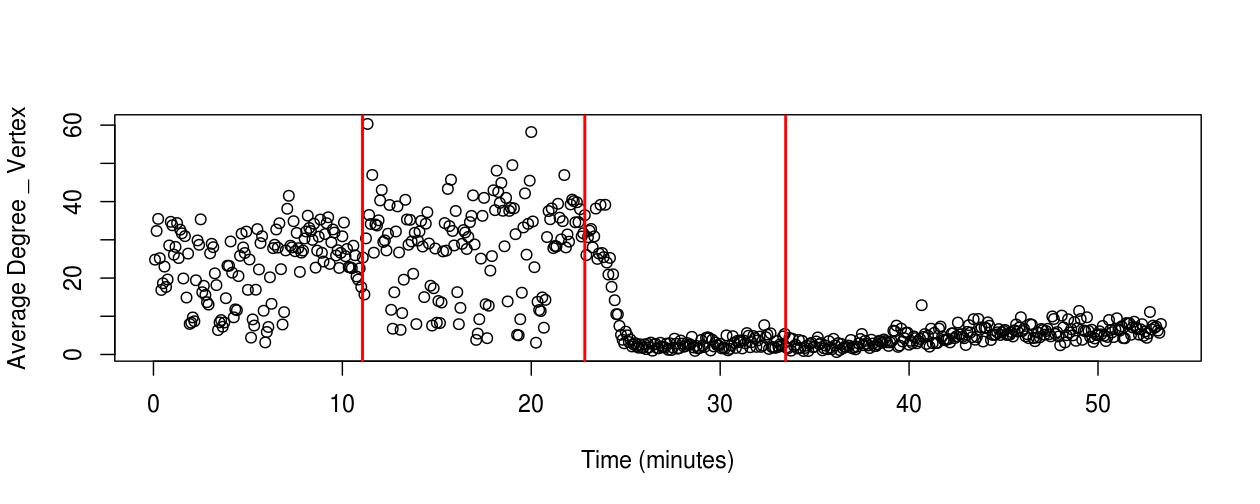}
  \caption{Beta 13-30Hz}
  \label{fig:sfig4}
\end{subfigure}\\
\centering
\begin{subfigure}{.5\textwidth}
\includegraphics[width=1\linewidth]{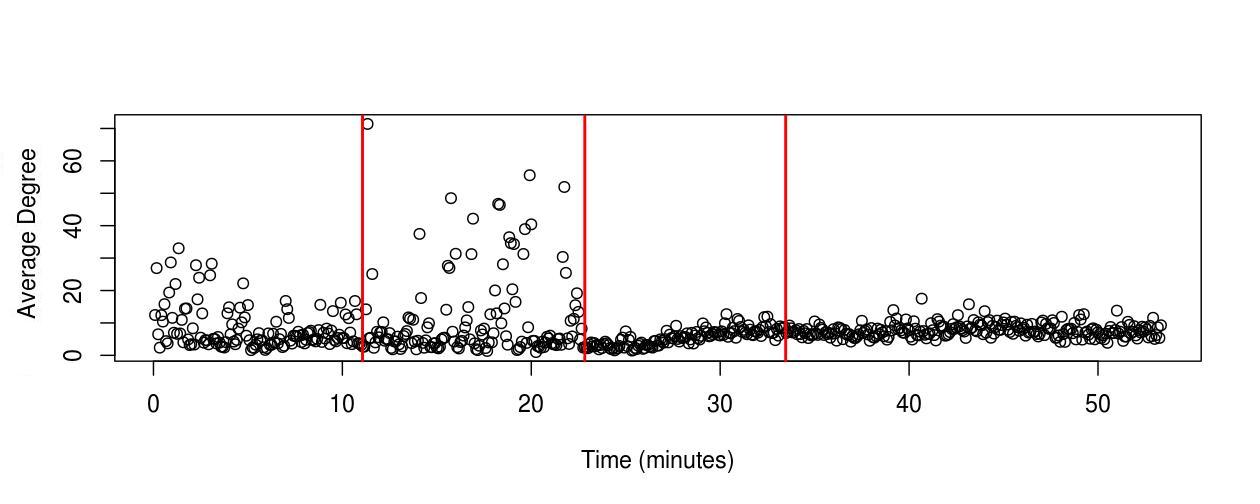}
  \caption{Gamma 25-100Hz}
  \label{fig:sfig3}
\end{subfigure}%
\begin{subfigure}{.5\textwidth}
  \centering
  \includegraphics[width=1\linewidth]{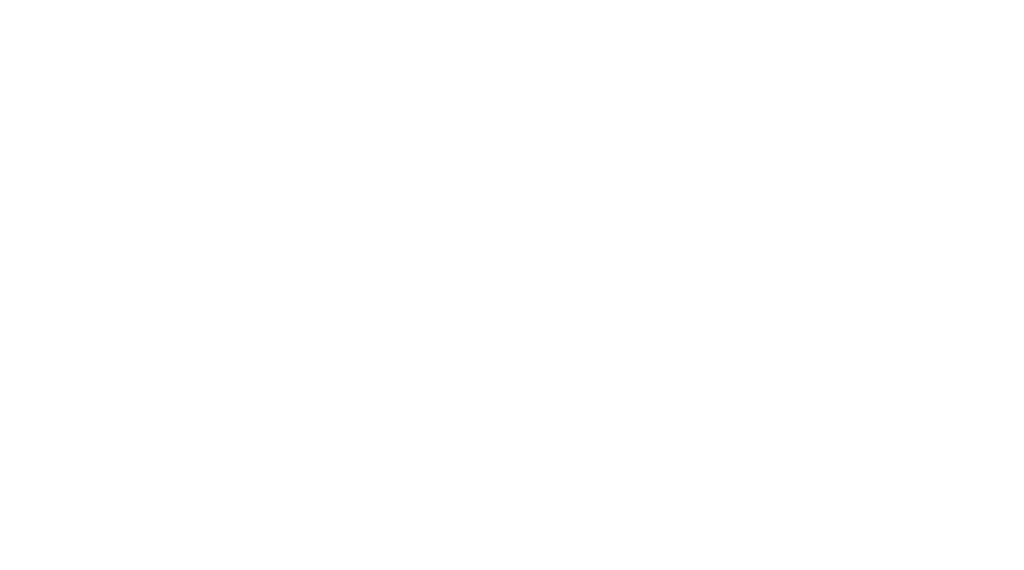}
  \label{fig:sfig4}
\end{subfigure}
\caption{\textbf{Average Degree Vertex}. Vertical axis average degree; Horizontal axis time (minutes). At t=11 minutes, the monkey was blindfolded; the first vertical red line represents this event in each sub-figure. At t=23 minutes, the Ketamine-Medetomidine cocktail was injected, represented by the second vertical red line. Finally, the point of loss of consciousness (LOC) was registered at t=33 minutes, indicated by the third vertical red line.
}
\hypertarget{FIGURE1}{}
\label{fig:fig}
\end{figure}

\begin{table}[!h]
\centering
\caption{Mean, variance (Var), and standard deviation (SD) of the average degree on the five physiological frequency bands analyzed and on the three different conditions in which the monkey was exposed during the experiment: awake with eyes open, awake with eyes closed and anesthesia (eyes closed). }
\vspace{0.5cm}
\begin{tabular}{l|lcr|lcr|lcr}
 \hline
\textbf{Average Degree} & \multicolumn{3}{c}{Eyes Open} \vline &\multicolumn{3}{c}{Eyes Closed} \vline &\multicolumn{3}{c}{Anesthesia}\\
\hline
Frequency Band & Mean & Var & SD  & Mean & Var & SD & Mean & Var & SD\\ 
\hline                             
Delta (0-4Hz)   &9.25   &30.3  &5.50    &17.4  &251   &15.8     &17.8    &121    & 11.0  \\
Theta  (4-8Hz)  &11.9   &28.6  &5.35    &29.3  &231   &15.2     &9.67    &16.0    &4.00   \\
Alpha (8-12Hz)  &10.6   &26.1  &5.11    &26.7  &189   &13.7     &3.21    &1.64    &1.28   \\
Beta (13-30Hz)  &23.6   &83.2  &9.12    &28.0  &152   &12.3     &4.47    &5.00    &2.23    \\
Gamma (25-100Hz)&8.29   &43.4  &6.59    &11.2  &163   &12.8     &7.64    &4.13    &2.03       %
\end{tabular}
\hypertarget{TABLE1}{}
\end{table}

\begin{multicols}{2}

\enlargethispage{1.5\baselineskip}
\vspace{-0.9\baselineskip}
\subsubsection*{Low Frequencies (0-4Hz)}

After the placement of the blindfold, the average degree respective to the Delta frequency band assumed a higher variation and approximately doubled in mean values (see \hyperlink{FIGURE1}{$Figure \cdot 1$, \mbox{Sub-Figure A}}; \hyperlink{TABLE1}{$Table \cdot 1$}). After administering the anesthetics, the degree of the graphs diminished and remained approximately constant for about four minutes. Later, an increase and a higher variation that lasted around 10 minutes were observed. After that time, the average degree fell again, keeping approximately constant until the end of the experiment. The networks respective to the awake (blindfolded) and anesthetized states had approximately the same mean connectivity (see \hyperlink{TABLE1}{$Table \cdot 1$}).

\subsubsection*{Medium Frequencies (4-30Hz)}

The average degree on the Theta and Alpha bands had increased significantly after the placement of the blindfold, having assumed a higher variation and approximately tripled in mean values (see \hyperlink{TABLE1}{$Table \cdot 1$}). In the Beta band, a similar dynamic behavior was also observed. In this frequency band, the graph's average connectivity increased by the order of 20\% due to the placement of the blindfold (see \hyperlink{TABLE1}{$Table \cdot 1$}).
 
After administering the anesthetics, the Theta, Alpha, and Beta bands presented a considerable reduction in the average degree compared to the time when the macaque was awake (blindfolded) (see \hyperlink{FIGURE1}{$Figure \cdot 1$, \mbox{Sub-Figures B, C, and D)}}, the average degree was reduced about three times in Theta, eight times in Alpha, and six times in the Beta band (see \hyperlink{TABLE1}{$Table \cdot 1$}). Under those experimental conditions, the variation of the graph's connectivity had diminished, and the networks assumed a quite constant degree along with the anesthetized state.

\subsubsection*{High Frequencies (25-100Hz)}

Between 25 and 100Hz, after covering the eyes of the monkey, the average degree increased in the order of 35\%. The connectivity of the networks assumed a higher variation, resulting in a two-fold increase in the standard deviation compared to the time when the macaque eyes were open (see \hyperlink{TABLE1}{$Table \cdot 1$}). After administering the anesthetics, the variation of the values decreased, and the average connectivity was reduced by approximately 30\% compared to the time when the macaque was awake (eyes closed) (see \hyperlink{TABLE1}{$Table \cdot 1$}).

\subsection{Correspondence Between Vertices Degree and Cortical Areas }

 \vspace{1\baselineskip}

Through the use of a color gradient, it was possible to verify the relationship between the degree of the vertices of the functional networks and their corresponding cortical anatomical areas. It was noted that most of the nodes that were physically closer to each other tended to present a similar degree, as they reflected in a similar color. However, it was also noticed that not all nodes had the same degree. The observed differences appeared to be related to anatomical areas and divisions (see \hyperlink{FIGURE2}{$Figure \cdot 2$}).
 
By analyzing the networks estimated serially over time, it was possible to observe changes among the patterns from one network to another, revealing that the connectivity of the vertices is dynamic. Besides the existence of some changes in consecutive (time) networks, two evident distinguishable patterns were observed along with the experiment. The first corresponds with the period when the monkey was awake, and the second pattern was observed when the macaque was anesthetized.

\subsubsection{Pattern of the Awake State}

The presence of high-degree nodes over the frontal and parietal lobes characterized the pattern observed in the awake state (see \hyperlink{FIGURE2}{$Figure \cdot 2$,        \mbox{Sub-Figures A - E)}}. Highly connected nodes were also observed, with a considerable likelihood on the medial frontal wall and anterior parts of the temporal lobe. It was also noted that there were areas where the degree of the vertices was relatively lower. Those areas encompassed mainly the occipital lobe, medial occipital walls, and medial-posterior temporal lobe areas.

\subsubsection{Pattern of the Anesthetized State}
 
The absence of high-degree nodes over the entire network characterized the pattern observed during the anesthesia-induced state. Despite the overall decrease in functional connectivity, at certain moments, some regions presented vertices with a higher degree. Those events of connectivity rise were mainly located in the occipital lobe and sometimes in the frontal and parietal areas (see \hyperlink{FIGURE2}{$Figure \cdot 2$, \mbox{Sub-Figures P - T)}}.

\end{multicols}

\begin{figure*}[!h]
  \includegraphics[width=\textwidth,height=15cm]
  {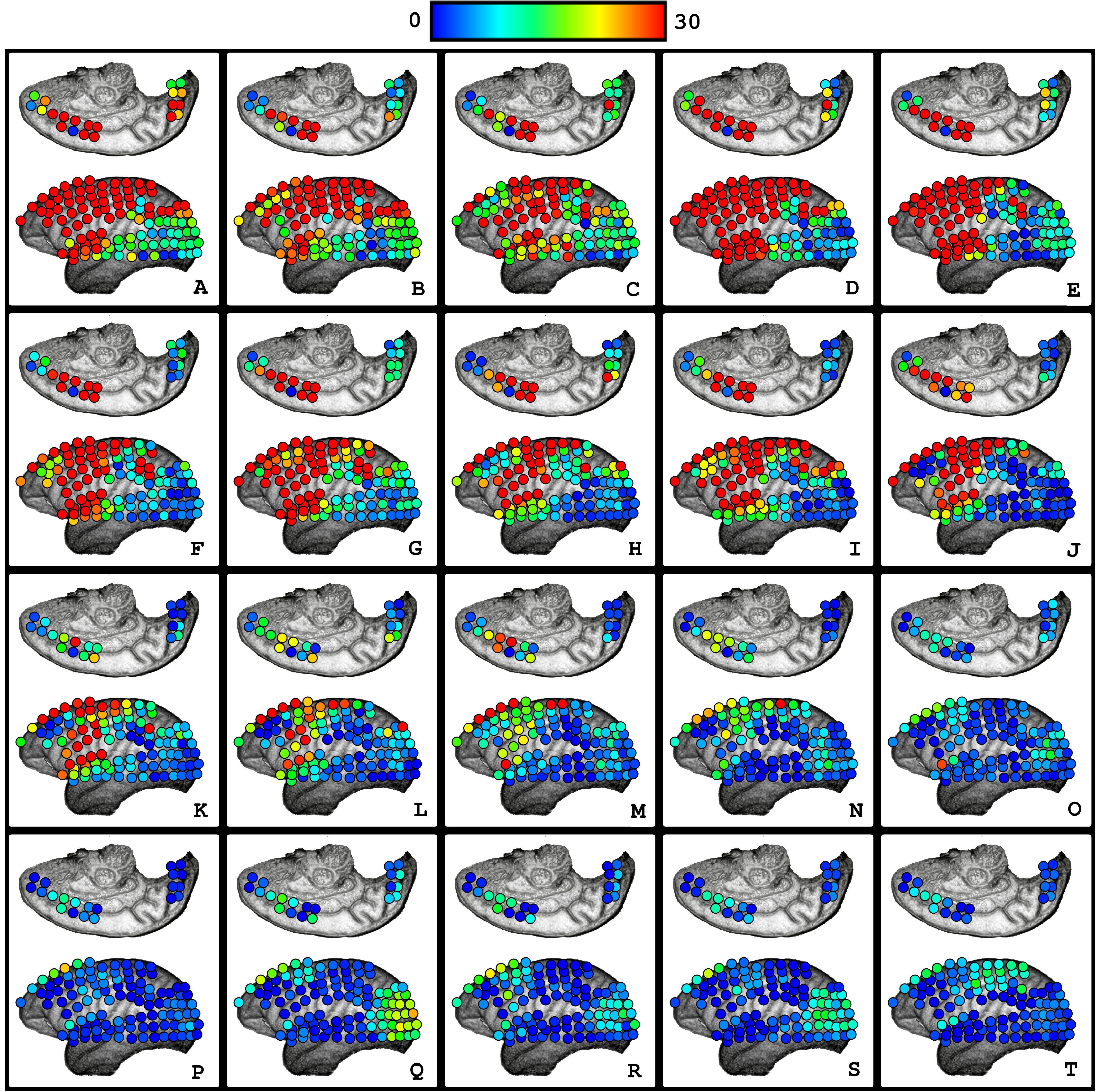}
     \caption{\textbf{Vertices Degree}. The transition between the awake and anesthetized states. Frequency band Alpha (8-12Hz). On each sub-figure, the vertex degree is indicated by a color gradient over the respective coordinates of each electrode. The sub-figures correspond to the degree of the vertices estimated sequentially over time, 5 seconds being the time interval between each frame and the subsequent one. The frame $A$ starts approximately one minute after administering the anesthetics.}
\hypertarget{FIGURE2}{}
\end{figure*}

\begin{multicols}{2}

\subsubsection*{Transition}

About one and a half minutes after the injection of the Ketamine-Medetomidine cocktail, an abrupt change in the degree of the vertices was observed, revealing a new and distinct pattern that persisted while the macaque was anesthetized. The transition between the two patterns occurred rapidly. From the observation of (\hyperlink{FIGURE2}{$Figure \cdot 2$}), it is possible to note that the transition took approximately 20-25 seconds \mbox{(4-5 networks)} (see \hyperlink{FIGURE2}{$Figure \cdot 2$}, \mbox{Sub-Figures  I - O).}

\vspace{12mm}

\subsection{Average Degree - Cortical Lobes}

Aimed at observing and comparing the alterations that occurred due to the anesthetic induction in each of the four cortical lobes, networks respective to each lobe were assembled (corresponding to subgraphs of the whole network).

\end{multicols}


\begin{figure}[!h]
\begin{subfigure}{.5\textwidth}
  \centering
  \includegraphics[width=1\linewidth]{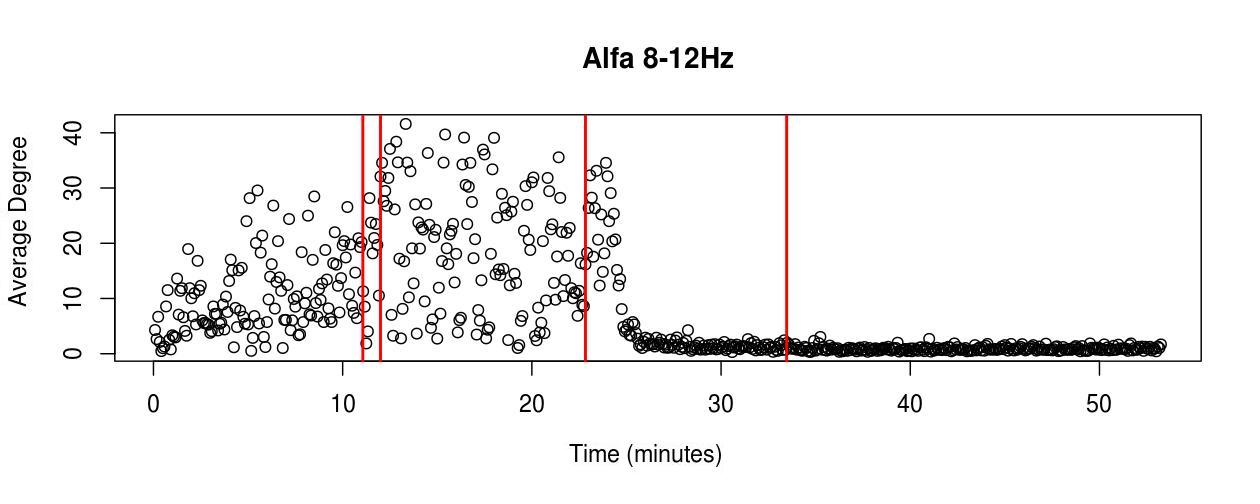}
  \caption{Frontal Lobe}
  \label{fig:sfig1}
\end{subfigure}%
\begin{subfigure}{.5\textwidth}
  \centering
  \includegraphics[width=1\linewidth]{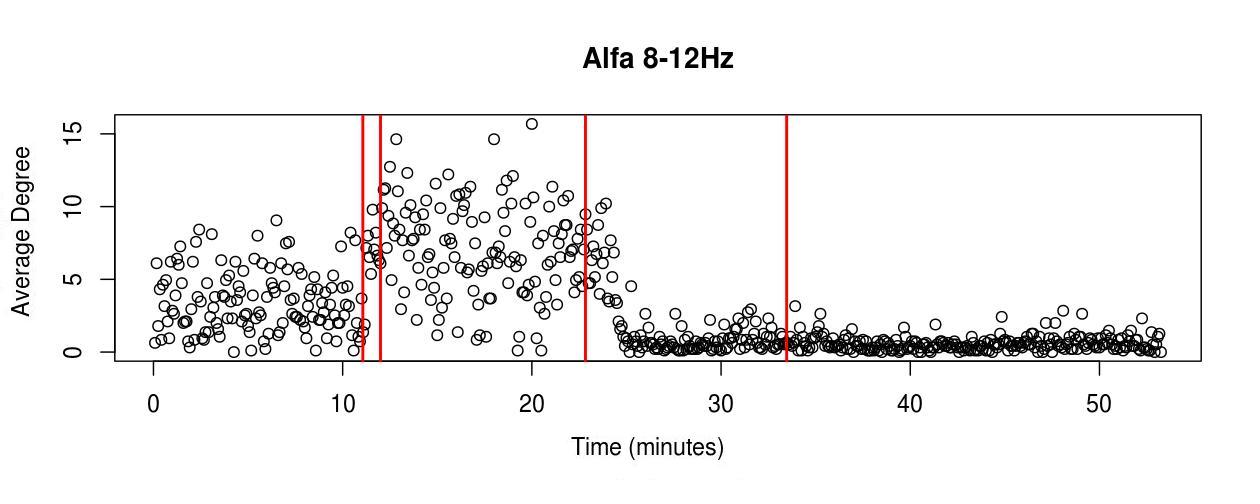}
  \caption{Parietal Lobe}
  \label{fig:sfig2}
\end{subfigure}\\
\centering
\begin{subfigure}{.5\textwidth}
\includegraphics[width=1\linewidth]{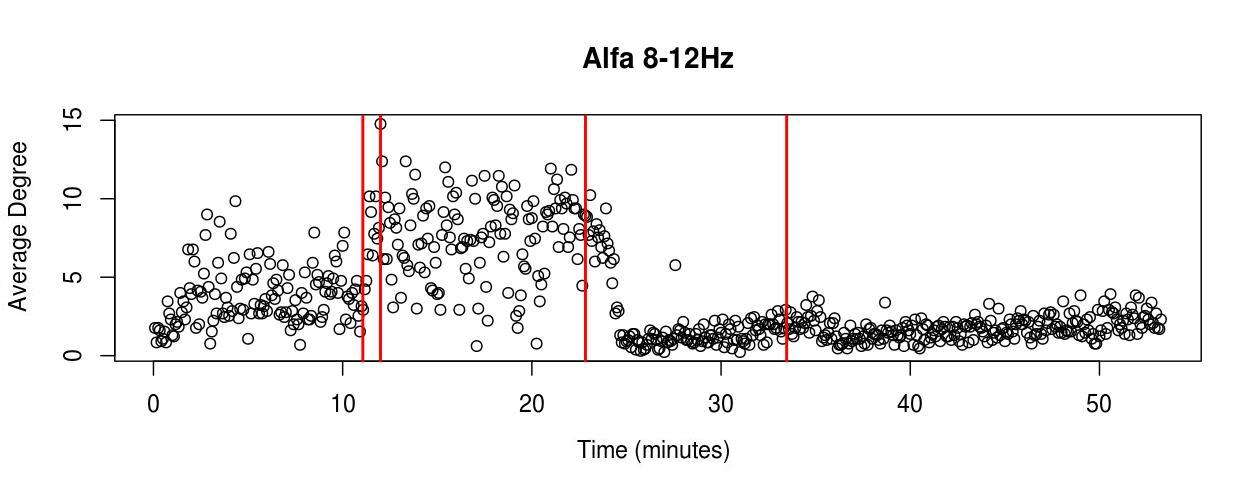}
  \caption{Temporal Lobe}
  \label{fig:sfig3}
\end{subfigure}%
\begin{subfigure}{.5\textwidth}
  \centering
  \includegraphics[width=1\linewidth]{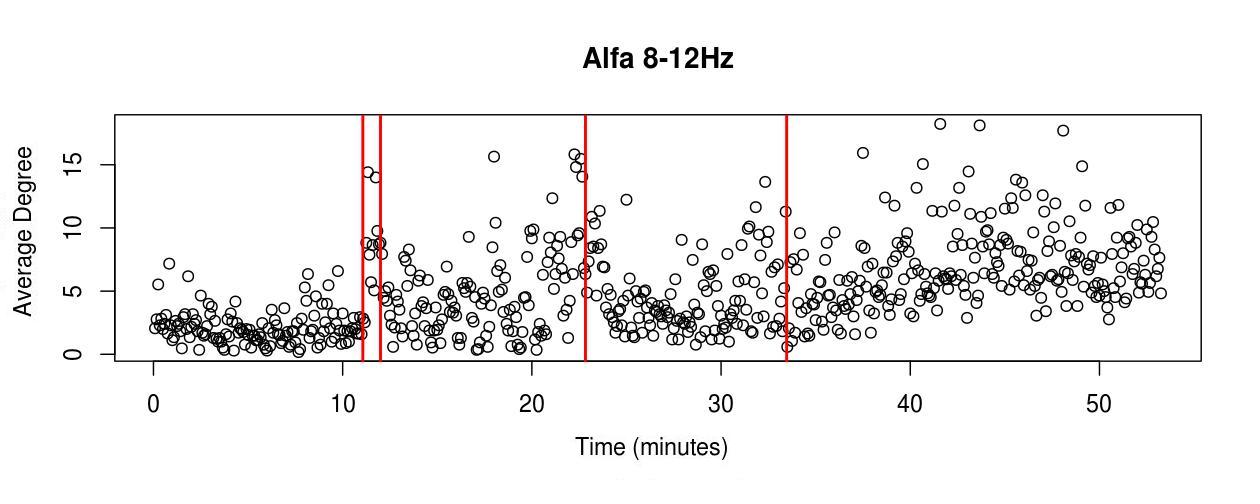}
  \caption{Occipital Lobe}
  \label{fig:sfig4}
\end{subfigure}
\caption{\textbf{Average Degree Vertex}. Vertical axis average degree in Alpha frequency band (8-12Hz); Horizontal axis time (minutes). At t=11 minutes, the monkey was blindfolded. The first two vertical red lines in each sub-figure represent the event of placing a patch over the eyes. At t=23 minutes, the third vertical red line represents the Ketamine-Medetomidine cocktail injection event. The point of loss of consciousness (LOC) was registered at t=33 minutes, indicated by the fourth vertical red line.}
\hypertarget{FIGURE3}{}
\label{fig:fig}
\end{figure}

\begin{table}[ht]
\centering
\caption{Mean, variance (Var), and standard deviation (SD) of the average degree of the complete graph (128 nodes), and the sub-graphs respective to each one of the four cortical lobes, on the frequency band Alpha (8-12Hz), on the three different conditions in which the monkey was exposed during the experiment: awake with eyes open, awake with eyes closed and anesthesia (eyes closed). }

\vspace{0.5cm}
\begin{tabular}{l|lcr|lcr|lcr}
\hline 
\textbf{Average Degree} & \multicolumn{3}{c}{Eyes Open} \vline &\multicolumn{3}{c}{Eyes Closed} \vline &\multicolumn{3}{c}{Anesthesia}\\
\hline
Corresponding Graph & Mean & Var & SD  & Mean & Var & SD & Mean & Var & SD\\ 
\hline                             
Complete Graph           &10.6     &26.1  &5.11   &26.7  &189   &13.75     &3.20   &1.64    &1.28   \\
Subgraph Frontal Lobe    &9.80     &44.3  &6.65   &18.0  &112   &10.6      &1.00    &0.24    &0.49   \\
Subgraph Parietal Lobe   &3.46     &4.41  &2.10   &6.82  &10.1   &3.17     &0.65    &0.32    &0.56   \\
Subgraph Temporal Lobe   &3.75     &3.57  &1.89   &7.52  &7.00   &2.65     &1.72    &0.51    &0.71    \\
Subgraph Occipital Lobe  &2.17     &1.96  &1.40   &4.80  &12.4   &3.52     &6.34    &10.7    &3.27       
 
\end{tabular}
\hypertarget{TABLE2}{}
\end{table}

\vspace{-0.5\baselineskip}

\begin{multicols}{2}

Significant alterations in the average degree at each one of the four brain lobes analyzed were verified over the experiment (see \hyperlink{FIGURE3}{$Figure \cdot 3$}; \hyperlink{TABLE2}{$Table \cdot 2$}).
 The frontal, parietal, and temporal regions presented quite similar dynamic behavior. In those regions, the mean connectivity degree increased and presented a higher variation after placing a blindfold over the eyes. A minute and a half after the injection of the anesthetic cocktail, an expressive decrease in the connectivity was noted (see \hyperlink{FIGURE3}{$Figure \cdot 3$, \mbox{Sub-Figures A, B, and C)}}. The average degree was reduced 18 times in the subgraphs of the frontal region, 10 times in the parietal lobe, and four times in temporal areas; in comparison to the awake state (blindfolded) (see \hyperlink{TABLE2}{$Table \cdot 2$}). In addition, a substantial decrease in the variation of the average degree on those cortical lobes was also observed, as the connectivity remained relatively constant during general anesthesia (see \hyperlink{FIGURE3}{$Figure \cdot 3$, \mbox{Sub-Figures A, B, and C).}}

\vspace{3mm}

A distinct dynamic behavior in response to the experimental conditions was observed on the sub-graphs respective to the occipital lobe.
After covering the eyes of the monkey, the average degree of the sub-graphs almost doubled and also presented a higher variation (see \hyperlink{FIGURE3}{$Figure \cdot 3$, \mbox{Sub-Figure D})}. After administering the anesthetics, a decrease in the mean connectivity lasting about seven minutes was verified. Then, the mean connectivity increased and showed a higher variation (see \hyperlink{FIGURE3}{$Figure \cdot 3$, \mbox{Sub-Figure D)}}.

\vspace{-0.5\baselineskip}

\subsection{Average Path Length}
\vspace{-0.4\baselineskip}

Noticeable changes in values of average path length were observed during the anesthetic induction process (see \hyperlink{FIGURE4}{$Figure \cdot 4$}; \hyperlink{TABLE2}{$Table \cdot 2$}).\\
 
Considering the effects of blindfolding the monkey, a subtle decrease in the average path length occurred. However, it was also possible to note that under this experimental condition, there was a tendency for the graphs to present quite the same values, keeping the average path length almost constant, except for short periods when higher values were observed.
 
Regarding the transition between the awake (blindfolded) and the anesthetized states, the anesthetic induction led to an expressive increase in the average path length of the networks. A higher variation of this property was verified during general anesthesia (see \hyperlink{FIGURE4}{$Figure \cdot 4$, \mbox{Sub-Figures B - E)}.}

\end{multicols}

\begin{figure}[!h]
\begin{subfigure}{.5\textwidth}
  \centering
  \includegraphics[width=1\linewidth]{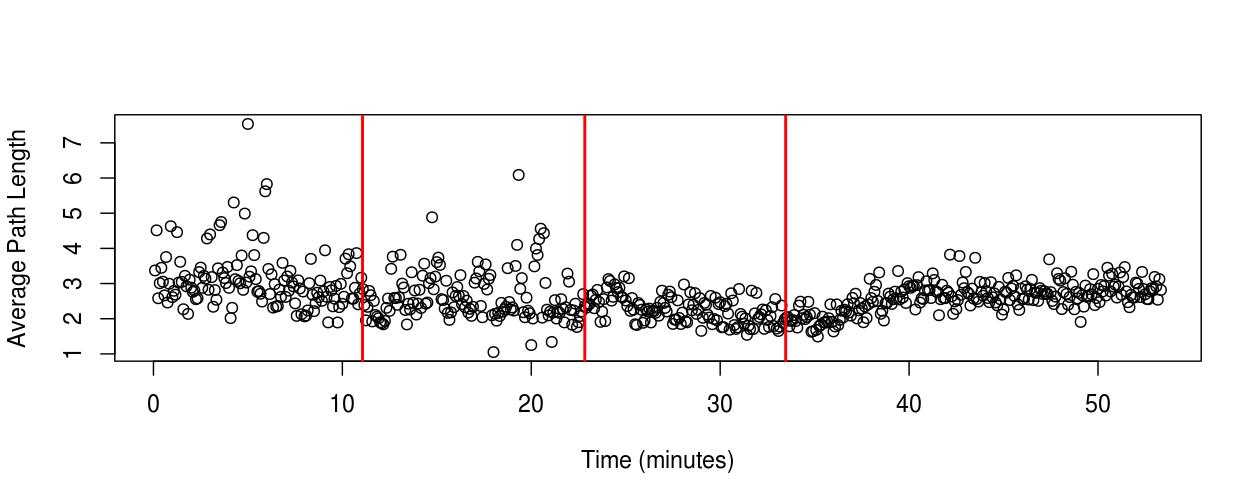}
  \caption{Delta 0-4Hz}
  \label{fig:sfig1}
\end{subfigure}%
\begin{subfigure}{.5\textwidth}
  \centering
  \includegraphics[width=1\linewidth]{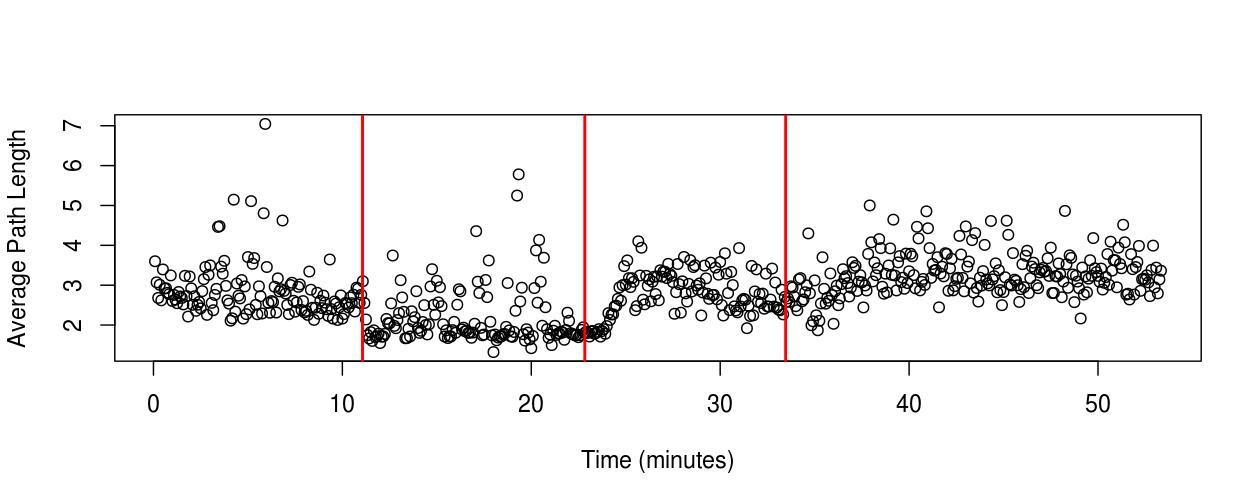}
 \caption{Theta 4-8Hz}
  \label{fig:sfig2}
\end{subfigure}\\
\centering
\begin{subfigure}{.5\textwidth}
\includegraphics[width=1\linewidth]{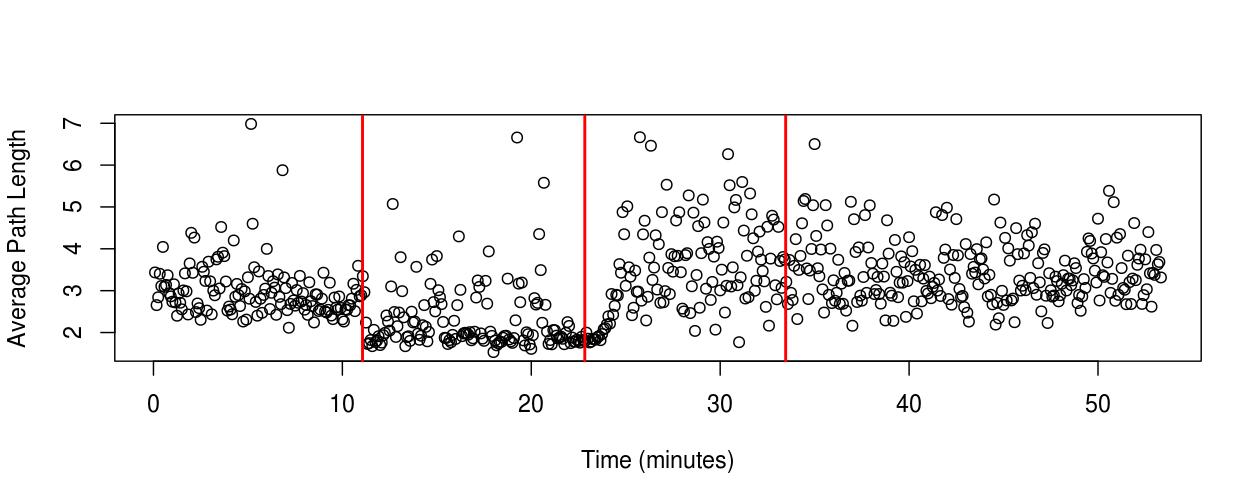}
  \caption{Alpha 8-12Hz}
  \label{fig:sfig3}
\end{subfigure}%
\begin{subfigure}{.5\textwidth}
  \centering
  \includegraphics[width=1\linewidth]{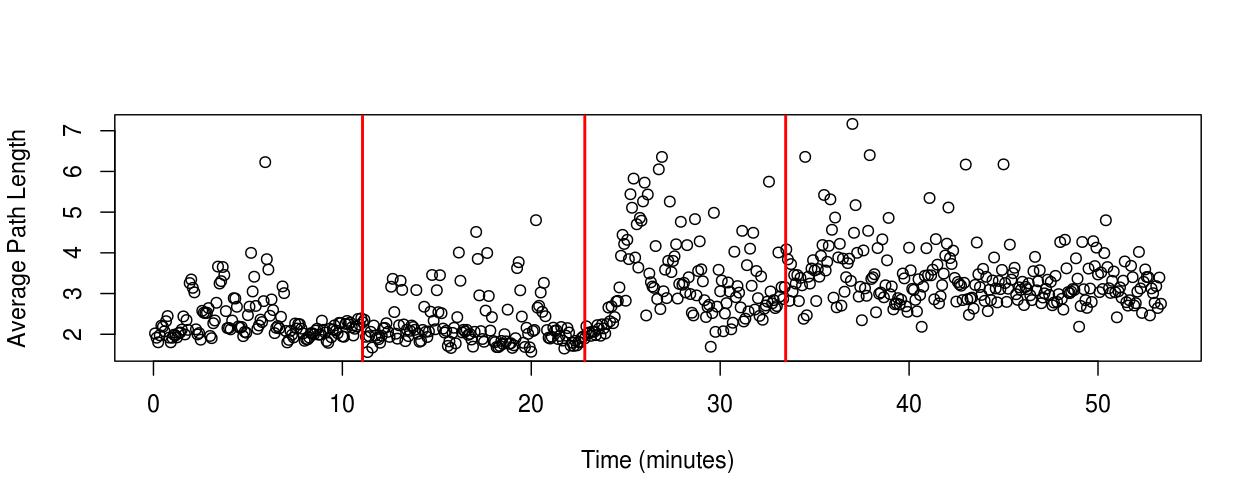}
 \caption{Beta 13-30Hz}
  \label{fig:sfig4}
\end{subfigure}\\
\centering
\begin{subfigure}{.5\textwidth}
\includegraphics[width=1\linewidth]{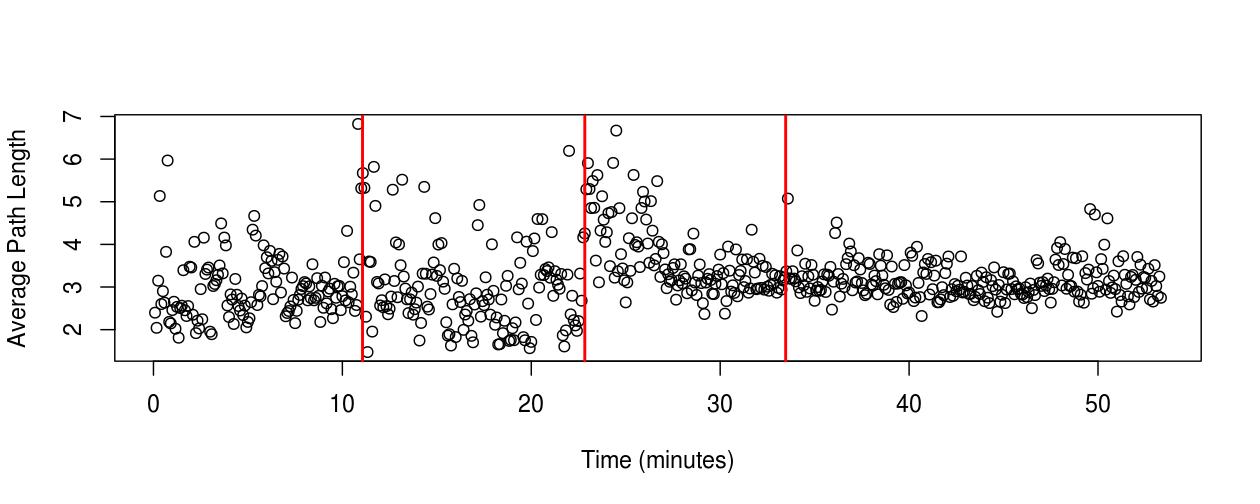}
 \caption{Gamma 25-100Hz}
  \label{fig:sfig3}
\end{subfigure}%
\begin{subfigure}{.5\textwidth}
  \centering
  \includegraphics[width=1\linewidth]{white}
  \label{fig:sfig4}
\end{subfigure}
\caption{\textbf{Average Path Length}. Vertical axis average path length; Horizontal axis time (minutes). At t=11 minutes, the monkey was blindfolded; the first vertical red line represents this event in each sub-figure. At t=23 minutes, the Ketamine-Medetomidine cocktail was injected, represented by the second vertical red line. Finally, the point of loss of consciousness (LOC) was registered at t=33 minutes, indicated by the third vertical red line.}
\label{fig:fig}
\hypertarget{FIGURE4}{}
\end{figure}

\begin{table}[h]
\centering
\caption{Mean, variance (Var), and standard deviation (SD) of the average path length on the five physiological frequency bands analyzed and on the three different conditions in which the monkey was exposed during the experiment: awake with eyes open, awake with eyes closed and anesthesia (eyes closed). }
\vspace{0.5cm}
\begin{tabular}{l|lcr|lcr|lcr}
 \hline
\textbf{Average Path} & \multicolumn{3}{c}{Eyes Open} \vline &\multicolumn{3}{c}{Eyes Closed} \vline &\multicolumn{3}{c}{Anesthesia}\\
\hline
Frequency Band & Mean & Var & SD  & Mean & Var & SD & Mean & Var & SD\\ %
\hline                             %
Delta (0-4Hz)   &3.11   &0.73  &0.85   &2.67  &0.56   &0.75     &2.50    &0.21    &0.46   \\
Theta  (4-8Hz)  &2.87   &0.50  &0.71   &2.20  &0.54   &0.73     &3.16    &0.30    &0.55   \\
Alpha (8-12Hz)  &3.03   &0.46  &0.68   &2.31  &0.64   &0.80     &3.51    &0.63    &0.79   \\
Beta (13-30Hz)  &2.37   &0.37  &0.61   &2.25  &0.40   &0.63     &3.34    &0.56    &0.75    \\
Gamma (25-100Hz)&2.94   &0.48  &0.69   &2.91  &0.81   &0.90     &3.16    &0.17    &0.41       
 
\end{tabular}
\hypertarget{TABLE3}{}
\end{table}

\begin{multicols}{2}

\subsubsection*{Low Frequencies (0-4Hz)}
\vspace{-2mm}

In the Delta band, no substantial changes in the average path length were observed along with the experiment (see \hyperlink{FIGURE4}{$Figure \cdot 4$, \mbox{Sub-Figure A)}}. Compared to when the monkey had its eyes open, blindfolding the macaque has led to a decrease in the order of 15\% in the average path length. After administering the anesthetics, a decrease in the order of 5\% was observed compared to the awake state (blindfolded) (see \hyperlink{TABLE3}{$Table \cdot 3$}).

\subsubsection*{Medium Frequencies (4-30Hz)}
 
On medium frequencies, the act of blindfolding the macaque led to a slight decrease in the average path length of the networks. Similar behavior occurred in the Theta, Alpha, and Beta bands during the anesthetic induction. A minute and a half after administering the Ketamine and Medetomidine cocktail, a substantial increase in the average path length in the order of 45\% in Theta and 50\% in Alpha and Beta occurred (see \hyperlink{TABLE3}{$Table \cdot 3$}; \hyperlink{FIGURE4}{$Figure \cdot 4$, \mbox{Sub-Figures B - D)}}. A higher variation in the average path length was also observed when the monkey was anesthetized.

\subsubsection*{High Frequencies (25-100Hz)}

On the Gamma frequency band, putting a patch over the eyes of the macaque has led to a higher variation in the average path length of the networks. After the anesthetic injection, the average path length increased. Four minutes later, some reduction was observed in the values. A decrease in the variation over time of the average path length was also verified during general anesthesia (see \hyperlink{FIGURE4}{$Figure \cdot 4$, \mbox{Sub-Figure E)}}.
  
  
\subsection{Diameter}

 Changes in the diameter of the networks were observed along with the experiment (see \hyperlink{FIGURE5}{$Figure \cdot 5$}; \hyperlink{TABLE4}{$Table \cdot 4$}).

\vspace{1.8\baselineskip}

The diameter of the graphs respective to the Delta band presented a distinct dynamic behavior compared to the other frequency bands. From 0 to 4Hz, the anesthetic induction reduced and decreased the diameter variation (see \hyperlink{FIGURE5}{$Figure \cdot 5$, \mbox{Sub-Figure A)}.}

\end{multicols}

\begin{figure}[!h]
\begin{subfigure}{.5\textwidth}
  \centering
  \includegraphics[width=1\linewidth]{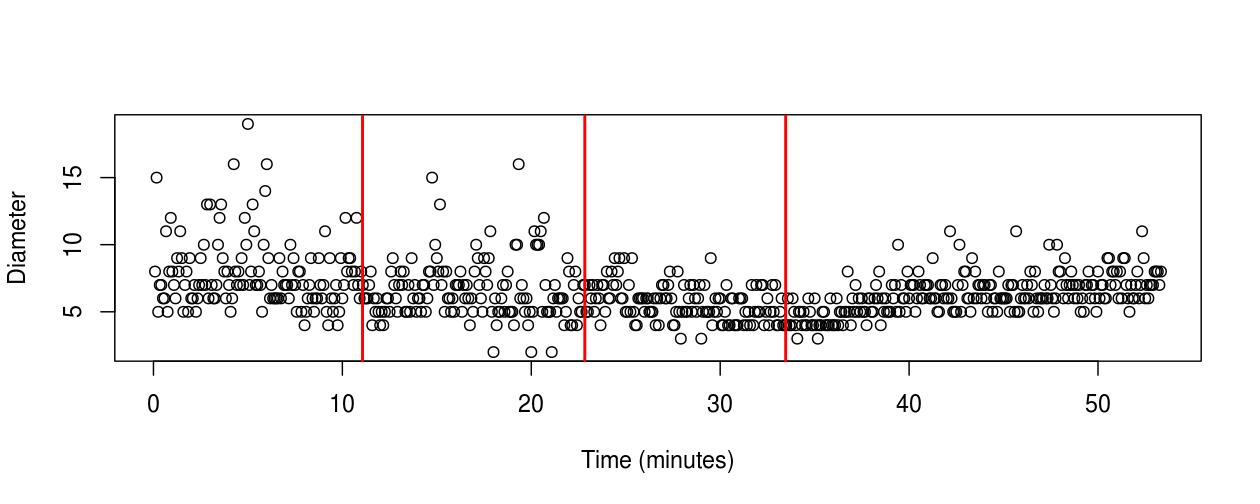}
  \caption{Delta 0-4Hz}
  \label{fig:sfig1}
\end{subfigure}%
\begin{subfigure}{.5\textwidth}
  \centering
  \includegraphics[width=1\linewidth]{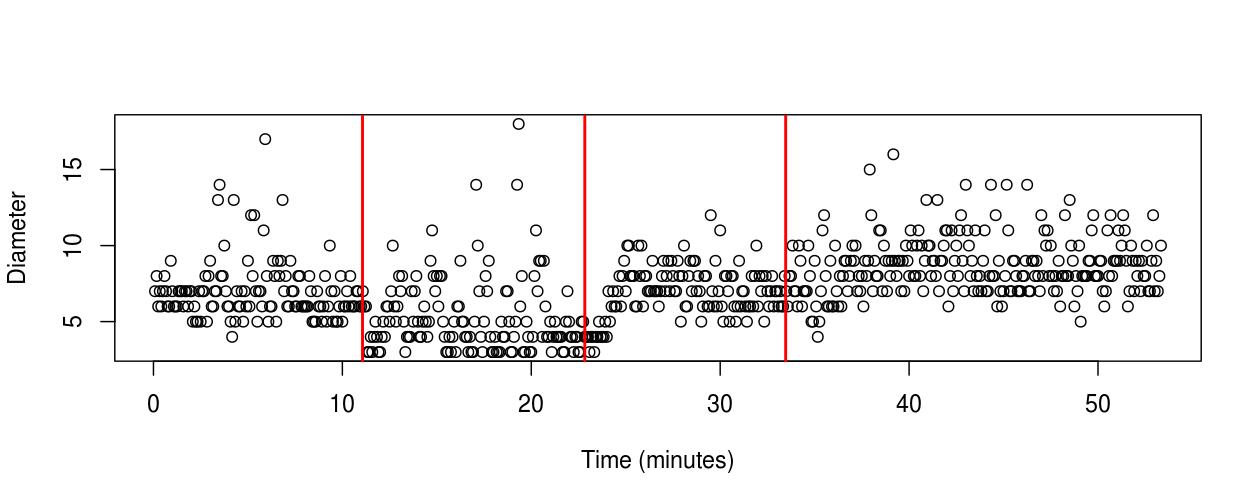}
 \caption{Theta 4-8Hz}
  \label{fig:sfig2}
\end{subfigure}\\
\centering
\begin{subfigure}{.5\textwidth}
\includegraphics[width=1\linewidth]{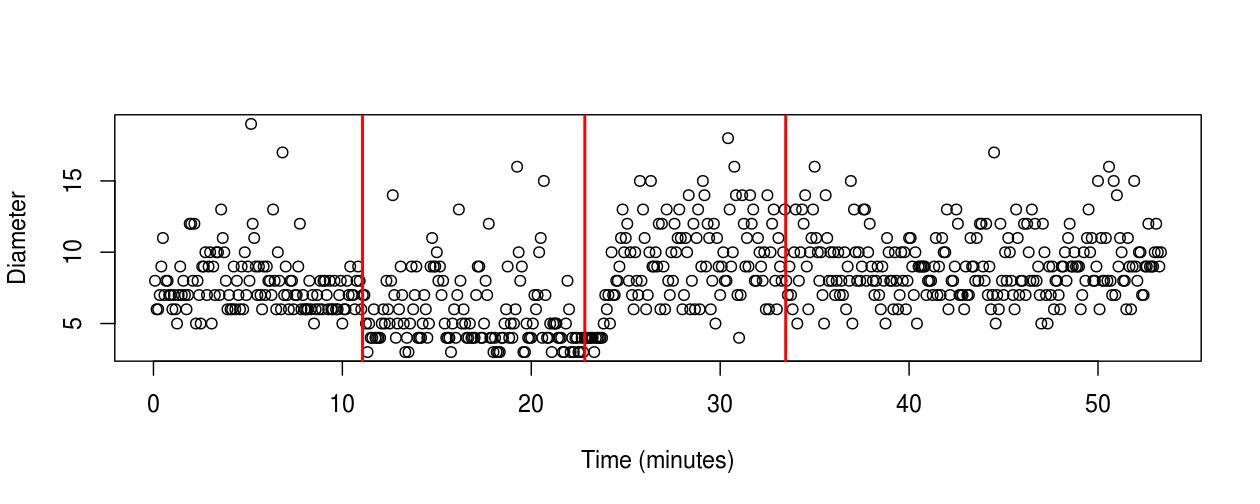}
  \caption{Alpha 8-12Hz}
  \label{fig:sfig3}
\end{subfigure}%
\begin{subfigure}{.5\textwidth}
  \centering
  \includegraphics[width=1\linewidth]{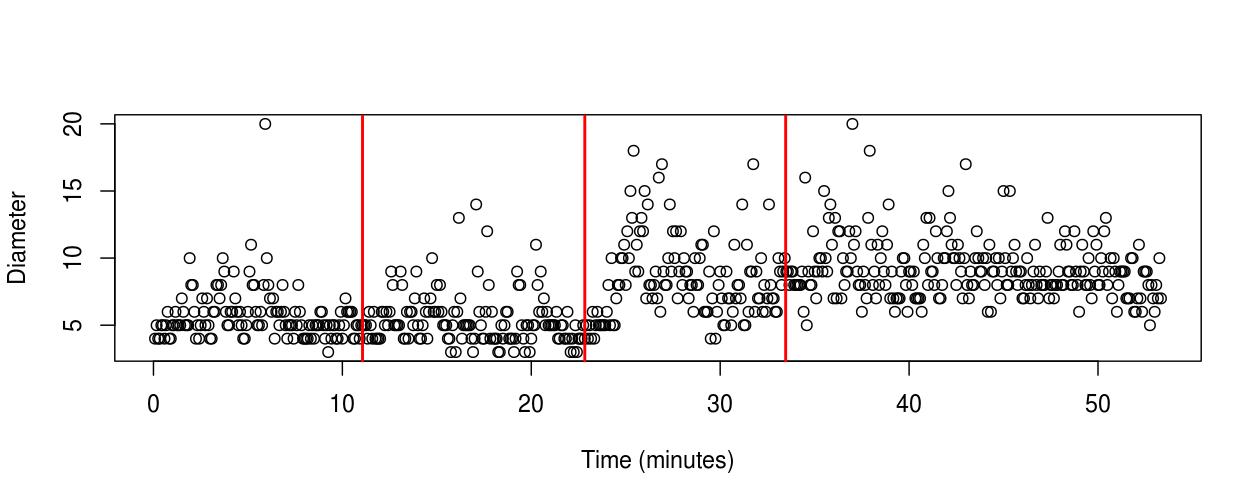}
 \caption{Beta 13-30Hz}
  \label{fig:sfig4}
\end{subfigure}\\
\centering
\begin{subfigure}{.5\textwidth}
\includegraphics[width=1\linewidth]{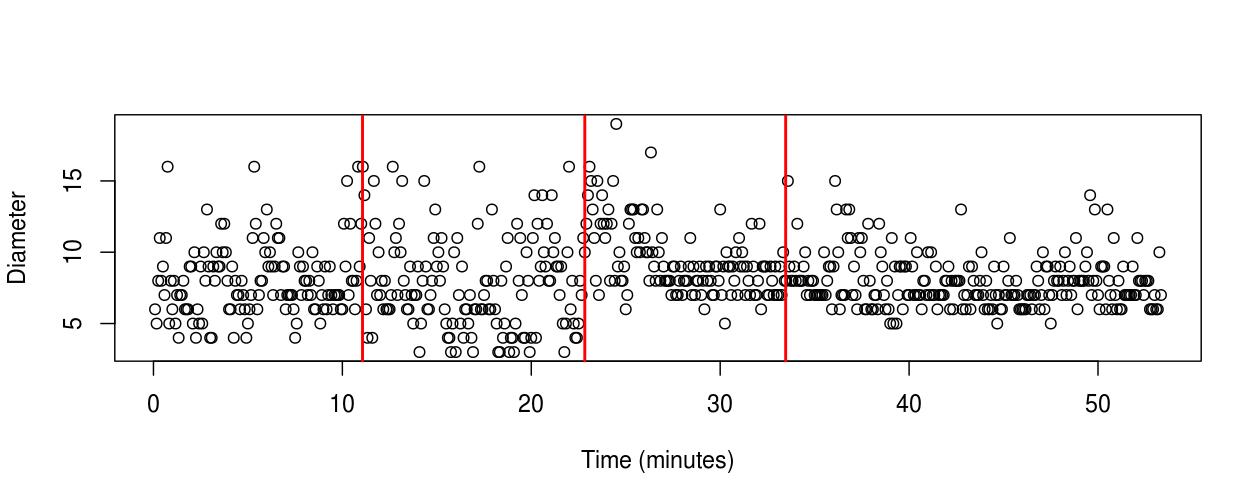}
\caption{Gamma 25-100Hz}
  \label{fig:sfig3}
\end{subfigure}%
\begin{subfigure}{.5\textwidth}
  \centering
  \includegraphics[width=1\linewidth]{white}
  \label{fig:sfig4}
\end{subfigure}
\caption{\textbf{Diameter}. Vertical axis diameter; Horizontal axis time (minutes). At t=11 minutes, the monkey was blindfolded; the first vertical red line represents this event in each sub-figure. At t=23 minutes, the Ketamine-Medetomidine cocktail was injected, represented by the second vertical red line. Finally, the point of loss of consciousness (LOC) was registered at t=33 minutes, indicated by the third vertical red line.}
\hypertarget{FIGURE5}{}
\label{fig:fig}
\end{figure}

\begin{table}[h]
\centering
\caption{Mean, variance (Var), and standard deviation (SD) of the diameter on the five physiological frequency bands analyzed and on the three different conditions in which the monkey was exposed during the experiment: awake with eyes open, awake with eyes closed, and anesthesia (eyes closed).}
\vspace{0.5cm}
\begin{tabular}{l|lcr|lcr|lcr}
\hline 
\textbf{Diameter} & \multicolumn{3}{c}{Eyes Open} \vline &\multicolumn{3}{c}{Eyes Closed} \vline &\multicolumn{3}{c}{Anesthesia}\\
\hline
Frequency Band & Mean & Var & SD  & Mean & Var & SD & Mean & Var & SD\\ 
\hline                             
Delta (0-4Hz)   &7.77     &7.15  &2.67   &6.70  &5.36   &2.31     &6,02    &2.12    &1.46   \\
Theta  (4-8Hz)  &7.05     &4.27  &2.07   &5.36  &6.62   &2.57     &8.28    &3.77    &1.94   \\
Alpha (8-12Hz)  &7.86     &5.03  &2.24   &5.67  &6.83   &2.61     &8.25    &6.21    &2.49   \\
Beta (13-30Hz)  &5.76     &4.16  &2.04   &5.42  &4.19   &2.05     &8.90    &5.38    &2.31    \\
Gamma (25-100Hz)&7.80     &5.44  &2.33   &7.60  &10.10  &3.18     &8.00    &3.10    &1.76       
 
\end{tabular}
\hypertarget{TABLE4}{}
\end{table}

\begin{multicols}{2}

Similar behavior was observed on the Theta, Alpha, Beta, and Gamma bands. After blindfolding the macaque, the diameter length subtly decreased, and a slight increase in the variation amplitude of the values was noted. One minute after the anesthetic cocktail administration, the graph's diameters increased substantially, staying this way until the end of the experiment (see \hyperlink{FIGURE5}{$Figure \cdot 5$, \mbox{Sub-Figures B - E)}}.

\subsection{ Average Betweenness Centrality Degree}
 \vspace{-0.4\baselineskip}
 
On all frequency bands analyzed, no significant changes in the average betweenness centrality degree of the vertices occurred after blindfolding the monkey (see \hyperlink{FIGURE6}{$Figure \cdot 6$}). Nonetheless, the administration of the anesthetics promoted changes in this property of the networks.
 
In the Delta band, administering the anesthetics decreased the vertex's average betweenness centrality degree, which lasted for about 15 minutes. Then, the values started increasing up to the recording section's end (see \hyperlink{FIGURE6}{$Figure \cdot 6$, \mbox{Sub-Figure A).}}


\end{multicols}

\begin{figure}[!h]
\begin{subfigure}{.5\textwidth}
  \centering
  \includegraphics[width=1\linewidth]{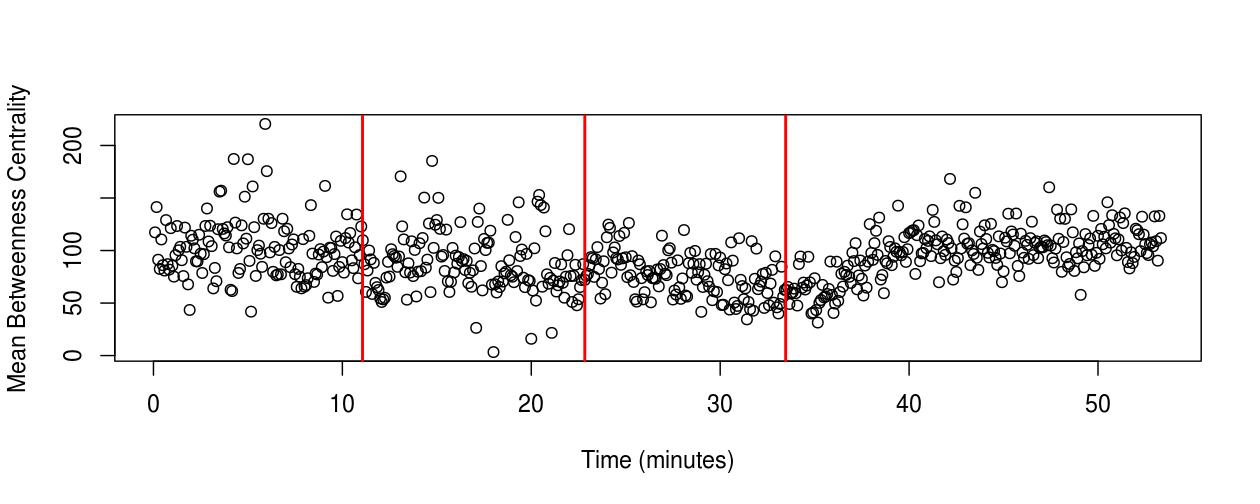}
  \caption{Delta 0-4Hz}
  \label{fig:sfig1}
\end{subfigure}%
\begin{subfigure}{.5\textwidth}
  \centering
  \includegraphics[width=1\linewidth]{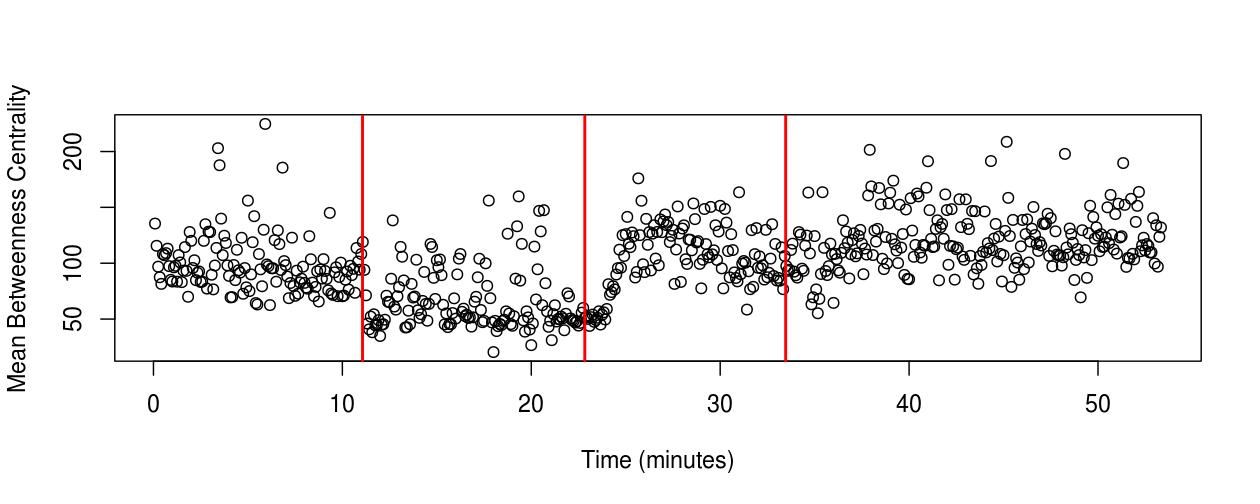}
  \caption{Theta 4-8Hz}
  \label{fig:sfig2}
\end{subfigure}\\
\centering
\begin{subfigure}{.5\textwidth}
\includegraphics[width=1\linewidth]{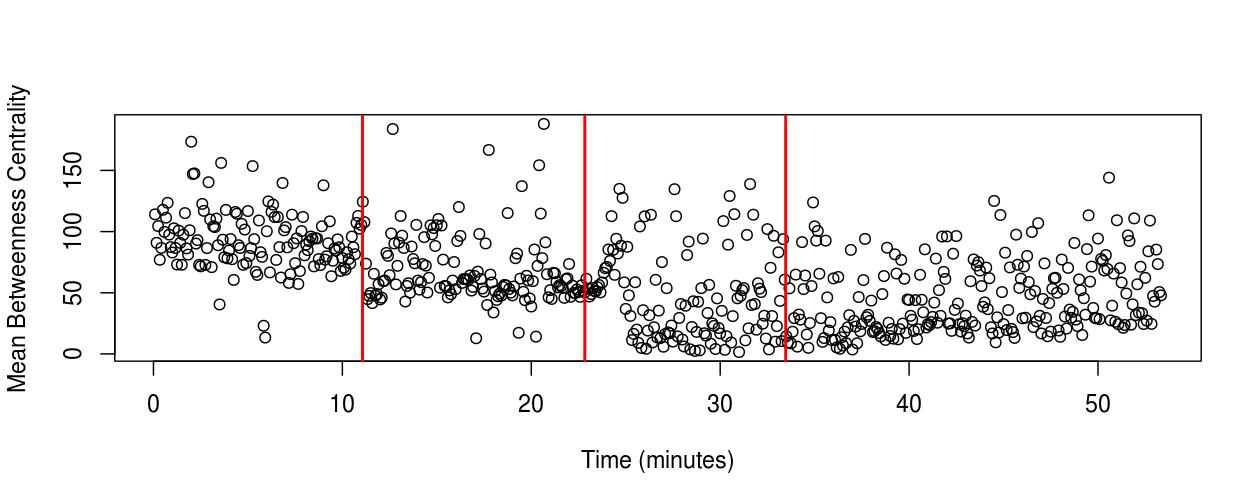}
  \caption{Alpha 8-12Hz}
  \label{fig:sfig3}
\end{subfigure}%
\begin{subfigure}{.5\textwidth}
  \centering
  \includegraphics[width=1\linewidth]{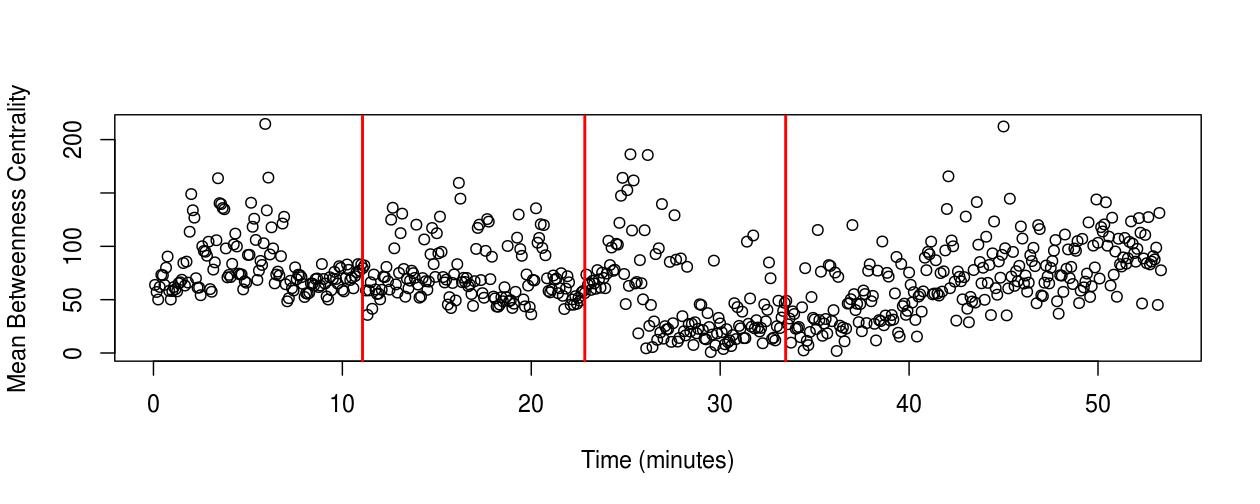}
 \caption{Beta 13-30Hz}
  \label{fig:sfig4}
\end{subfigure}\\
\centering
\begin{subfigure}{.5\textwidth}
\includegraphics[width=1\linewidth]{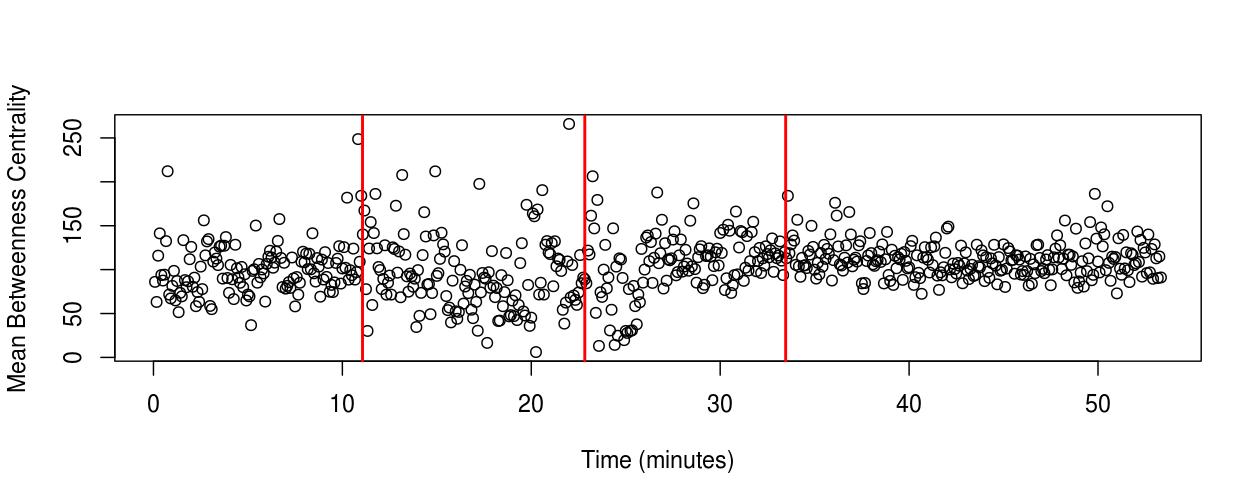}
\caption{Gamma 25-100Hz}
  \label{fig:sfig3}
\end{subfigure}%
\begin{subfigure}{.5\textwidth}
  \centering
  \includegraphics[width=1\linewidth]{white}
 
  \label{fig:sfig4}
\end{subfigure}
\caption{
\textbf{Betweenness Centrality}. Vertical axis betweenness centrality; Horizontal axis time (minutes). At t=11 minutes, the monkey was blindfolded; the first vertical red line represents this event in each sub-figure. At t=23 minutes, the Ketamine-Medetomidine cocktail was injected, represented by the second vertical red line. Finally, the point of loss of consciousness (LOC) was registered at t=33 minutes, indicated by the third vertical red line.
}
\hypertarget{FIGURE6}{}
\label{fig:fig}
\end{figure}

\begin{table}[h]
\centering
\caption{Mean, variance (Var), and standard deviation (SD) of the mean betweenness centrality degree on the five physiological frequency bands analyzed and on the three different conditions in which the monkey was exposed during the experiment: awake with eyes open, awake with eyes closed and anesthesia (eyes closed).}
\vspace{0.5cm}
\begin{tabular}{l|lcr|lcr|lcr}
\hline 
\textbf{Centrality} & \multicolumn{3}{c}{Eyes Open} \vline &\multicolumn{3}{c}{Eyes Closed} \vline &\multicolumn{3}{c}{Anesthesia}\\
\hline
Frequency Band & Mean & Var & SD  & Mean & Var & SD & Mean & Var & SD\\ 
\hline                             
Delta (0-4Hz)   &102      &875  &29.6   &88.6  &920   &30.3     &91.2    &662    &25.7   \\
Theta  (4-8Hz)  &99.5     &744  &27.3   &67.5  &822   &28.7     &117     &642    &25.3   \\
Alpha (8-12Hz)  &93.1     &599  &24.5   &67.5  &761   &27.6     &45.0    &932    &30.5   \\
Beta (13-30Hz)  &82.5     &837  &28.9   &73.3  &710   &26.6     &60.8    &1270   &35.6    \\
Gamma (25-100Hz)&98.2     &667  &25.8   &92.2  &1840  &42.9     &112     &413    &20.3      
 
\end{tabular}
\hypertarget{TABLE5}{}
\end{table}

\begin{multicols}{2}

In both Theta and Gamma bands, after the administration of the anesthetics, an increase in the average betweenness centrality degree and a reduction in the variation of the values were observed compared to the awake state (eyes open and blindfolded) (see \hyperlink{FIGURE6}{$Figure \cdot 6$, \mbox{Sub-Figures B and E)}}.
 
In the Alpha band, after administering the anesthetics, the vertice's average centrality degree was reduced, and an increase in the variation of the values was observed (see \hyperlink{FIGURE6}{$Figure \cdot 6$, \mbox{Sub-Figure C)}}.
 
On the Beta band, one and a half minutes after the anesthetic injection, the average betweenness centrality degree of the nodes was reduced. About 15 minutes later, the values increased (see \hyperlink{FIGURE6}{$Figure \cdot 6$, \mbox{Sub-Figure D)}}.


\subsection{Vertices Betweenness Centrality Degree and Cortical Areas}

By plotting the intermediation degree of each vertice through a color gradient, it was possible to verify the correspondence between the intermediation degree of the network's vertices and their corresponding cortical areas and anatomical divisions.

It was noticeable that nodes physically closer to each other presented a quite similar intermediation degree. However, not all vertices of the networks had the same betweenness centrality degree. Furthermore, it was possible to note that the betweenness centrality degree of the vertices was related to cortical anatomy and divisions (see \hyperlink{FIGURE7}{$Figure \cdot 7$}). 
 
It was verified that the intermediation degree of the vertices is dynamic once variations from consecutive (time) networks have occurred. Besides those variations, during the experiment, two prominent patterns were observed, the first respective to the time when the monkey was awake and the second to when it was anesthetized.
 
\end{multicols}

\begin{figure*}[!h]
  \includegraphics[width=\textwidth,height=15cm]
  {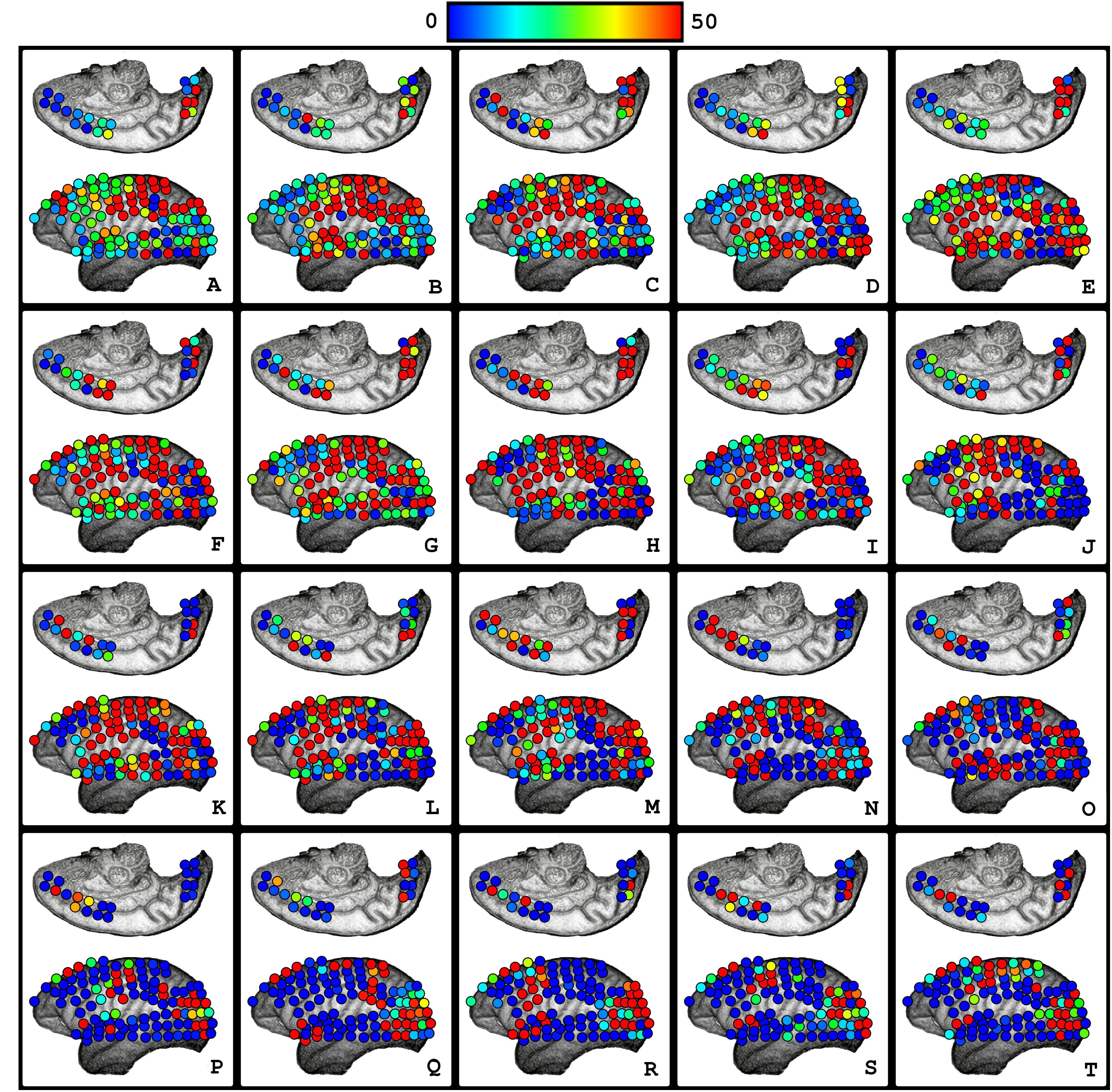}
     \caption{\textbf{Vertice's Intermediation Degree}. The transition between the awake and anesthetized states. Frequency band Alpha (8-12Hz). In each sub-figure, the vertice's intermediation degree is represented through a color gradient over the respective coordinates of the electrodes. The sub-figures correspond to the intermediation degree of the vertices estimated sequentially over time, being 5 seconds the time interval between each frame and its subsequent. The frame $A$ starts approximately one minute after the administration of the anesthetics.}
\hypertarget{FIGURE7}{}
\end{figure*}

\begin{multicols}{2}

\subsubsection*{Awake State Pattern}

While in awakening conditions, most nodes had a relevant contribution to the network's geodesic paths. Notably, some vertices had a higher intermediation degree and were frequently located closer to each other. Moreover, those higher-degree vertices extended and involved continuous large areas of the cortex (see \hyperlink{FIGURE7}{$Figure \cdot 7$, \mbox{Sub-Figures A - E)}}.

 \newpage
 
\subsubsection*{Anesthetized State Pattern}
 
While the macaque was anesthetized, a few nodes (cortical areas) had a high betweenness centrality degree, while the other network nodes had a reduced centrality degree. It was possible to observe that the vertices of some specific cortical regions monopolized the network's intermediation. Another remarkable change observed was the coverage discontinuity of those areas of high betweenness centrality located far from each other, separated by regions characterized by vertices of reduced intermediation degree (see \hyperlink{FIGURE7}{$Figure \cdot 7$, \mbox{Sub-Figures P - T)}}.

\subsubsection*{Transition}

About one minute and a half after administering the anesthetics, there was a transition in the patterns of the vertices intermediation degree. The pattern observed in the awake state was no longer present, being replaced by a different pattern that predominated while the monkey was anesthetized. The transition between the two patterns occurred in approximately 40 to 50 seconds (see \hyperlink{FIGURE7}{$Figure \cdot 7$, \mbox{Sub-Figures J - Q)}.}
 
\subsection{Assortativity}
 
The placement of a blindfold over the eyes of the monkey did not change the assortative coefficient of the graphs significantly (see \hyperlink{FIGURE8}{$Figure \cdot 8$}). Small changes were observed in each frequency band. However, no significant alteration in the dynamic behavior occurred as a result of blindfolding the macaque (eyes open vs. closed).


\end{multicols}

\begin{figure}[!h]
\begin{subfigure}{.5\textwidth}
  \centering
  \includegraphics[width=1\linewidth]{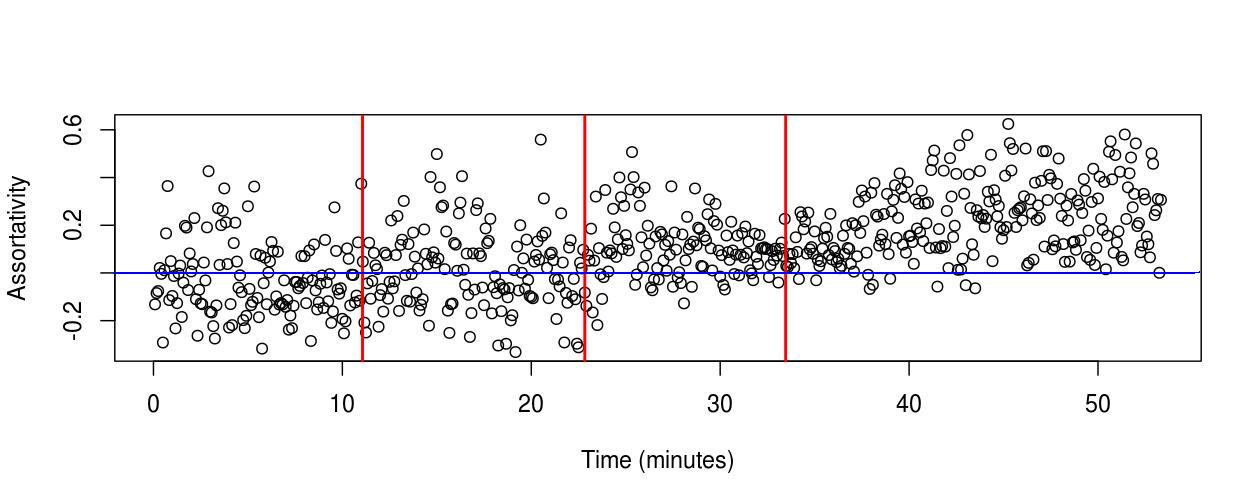}
  \caption{Delta 0-4Hz}
  \label{fig:sfig1}
\end{subfigure}%
\begin{subfigure}{.5\textwidth}
  \centering
  \includegraphics[width=1\linewidth]{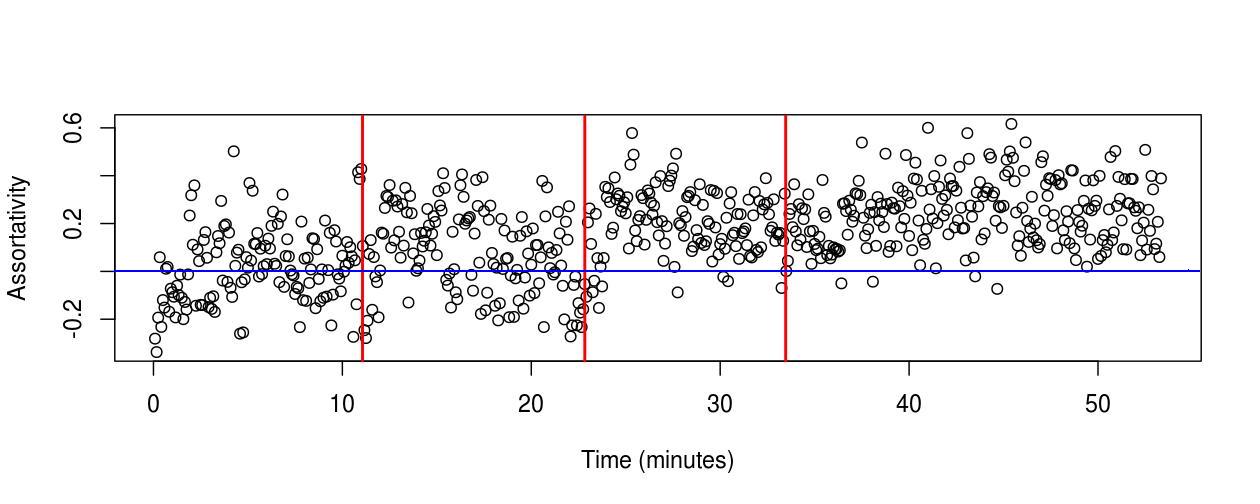}
 \caption{Theta 4-8Hz}
  \label{fig:sfig2}
\end{subfigure}\\
\centering
\begin{subfigure}{.5\textwidth}
\includegraphics[width=1\linewidth]{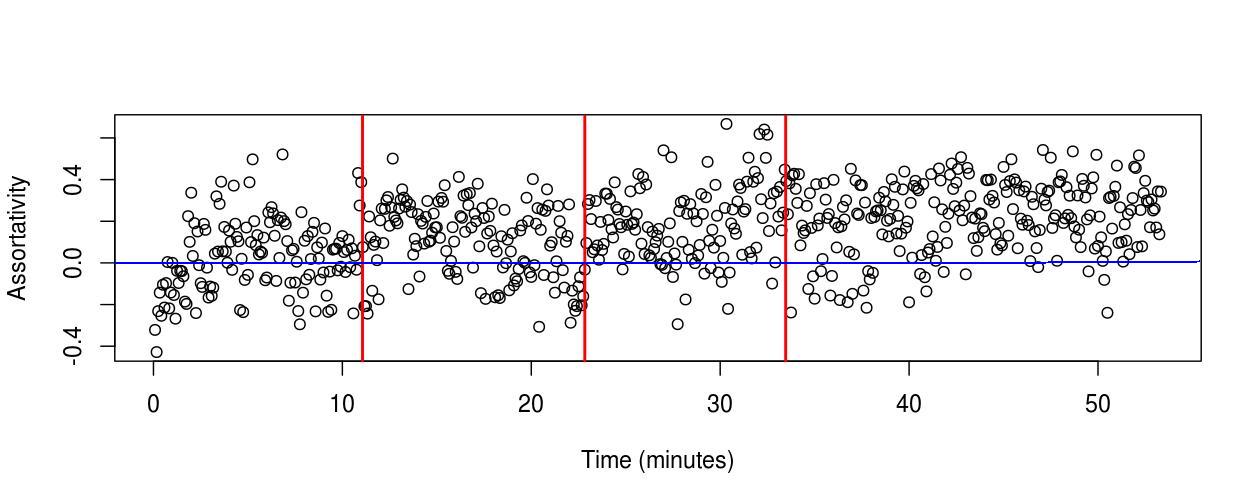}
  \caption{Alpha 8-12Hz}
  \label{fig:sfig3}
\end{subfigure}%
\begin{subfigure}{.5\textwidth}
  \centering
  \includegraphics[width=1\linewidth]{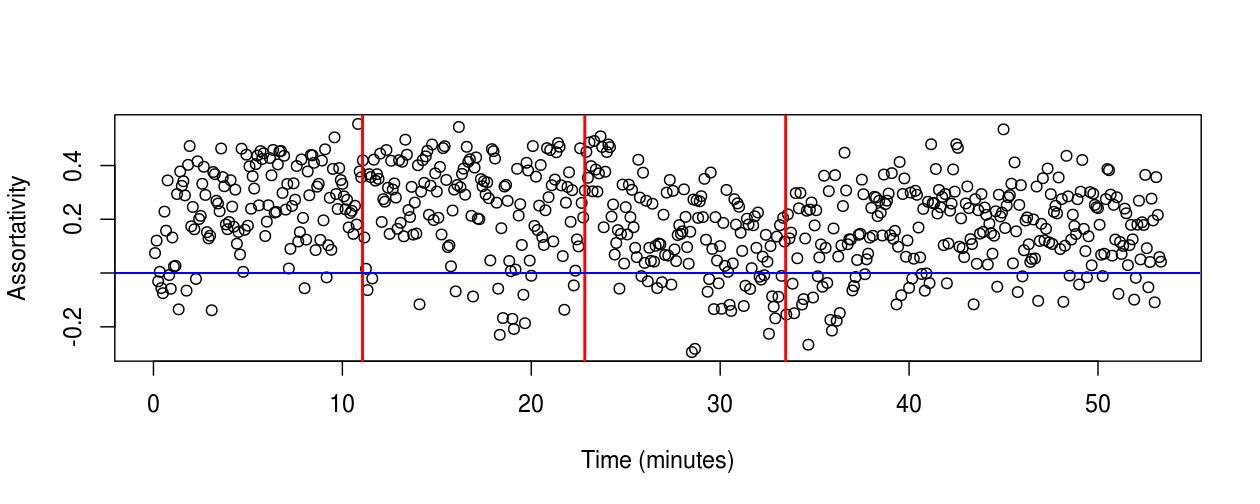}
 \caption{Beta 13-30Hz}
  \label{fig:sfig4}
\end{subfigure}\\
\centering
\begin{subfigure}{.5\textwidth}
\includegraphics[width=1\linewidth]{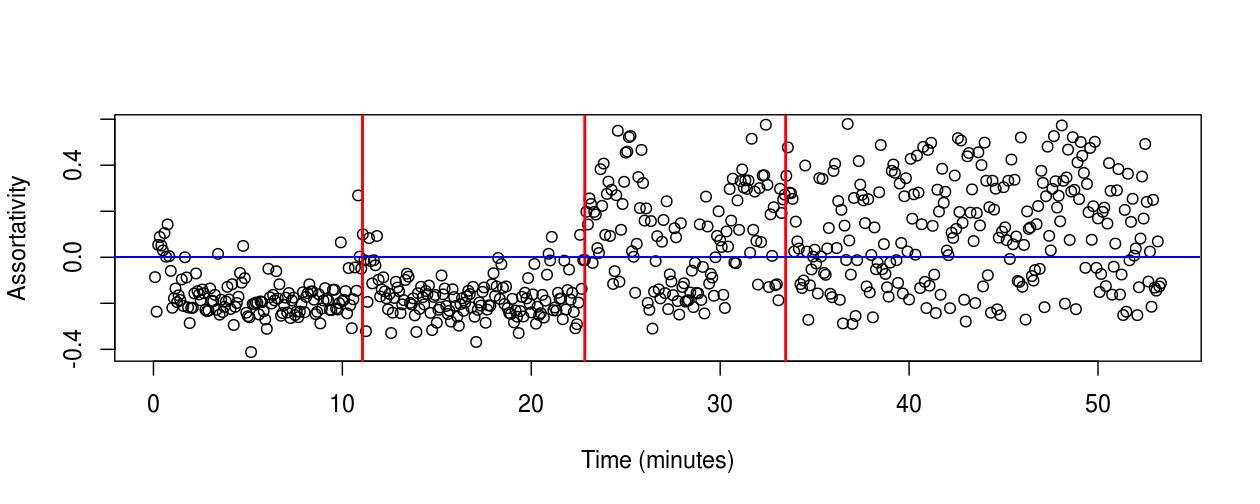}
\caption{Gamma 25-100Hz}
  \label{fig:sfig3}
\end{subfigure}%
\begin{subfigure}{.5\textwidth}
  \centering
  \includegraphics[width=1\linewidth]{white}
  \label{fig:sfig4}
\end{subfigure}
\caption{\textbf{Assortativity}. Vertical axis assortativity; Horizontal axis time (minutes). At t=11 minutes, the monkey was blindfolded; the first vertical red line represents this event in each sub-figure. Next, at t=23 minutes, the Ketamine-Medetomidine cocktail was injected, represented by the second vertical red line. Finally, the point of loss of consciousness (LOC) was registered at t=33 minutes, indicated by the third vertical red line.}
\label{fig:fig}
\hypertarget{FIGURE8}{}
\end{figure}

\begin{table}[h]
\centering
\caption{Mean, variance (Var), and standard deviation (SD) of the assortativity coefficient on the five physiological frequency bands analyzed and on the three different conditions in which the monkey was exposed during the experiment: awake with eyes open, awake with eyes closed and anesthesia (eyes closed). }
\vspace{0.5cm}
\begin{tabular}{l|lcr|lcr|lcr}
\hline 
\textbf{Assortativity} & \multicolumn{3}{c}{Eyes Open} \vline &\multicolumn{3}{c}{Eyes Closed} \vline &\multicolumn{3}{c}{Anesthesia}\\
\hline
Frequency Band & Mean & Var & SD  & Mean & Var & SD & Mean & Var & SD\\ 
\hline                             
Delta (0-4Hz)   &-0.03    &0.02    &0.14   &0.02   &0.03   &0.17     &0.19    &0.02    &0.14   \\
Theta  (4-8Hz)  &0.01     &0.02    &0.14   &0.08   &0.03   &0.17     &0.23    &0.02    &0.14   \\
Alpha (8-12Hz)  &0.02     &0.03    &0.17   &0.10   &0.03   &0.17     &0.22    &0.03    &0.17   \\
Beta (13-30Hz)  &0.25     &0.02    &0.14   &0.25   &0.03   &0.17     &0.14    &0.02    &0.14    \\
Gamma (25-100Hz)&-0.17    &0.01    &0.10   &-0.18  &0.01   &0.17     &0.11    &0.05    &0.22       
 
\end{tabular}
\hypertarget{TABLE6}{}
\end{table}

\begin{multicols}{2}

Significant changes in the network's assortative character occurred after administering the anesthetics (see \hyperlink{FIGURE8}{$Figure \cdot 8$}; \hyperlink{TABLE6}{$Table \cdot 6$}). Quite similar behavior was observed in the Delta, Theta, and Alpha bands. While the macaque was awake, the assortativity varied in those frequency bands, assuming positive and negative values. On average, a disassortative character was prevalent. A few seconds after the administration of the anesthetics, an accentuated change occurred, and the graphs assumed an assortative character, having positive assortativity (see \hyperlink{FIGURE8}{$Figure \cdot 8$, \mbox{Sub-Figures A, B, and C)}.}
 
\vspace{4.5\baselineskip}
 
In the Beta band, an assortative character prevailed throughout the experiment. After the anesthetics were administered, the assortativity gradually lowered for about 10 minutes. After that time, the assortativity started to increase and assumed the same dynamic behavior until the end of the experiment (see \hyperlink{FIGURE8}{$Figure \cdot 8$, \mbox{Sub-Figure D)}}.

While the monkey was awake, the networks respective to the Gamma band were disassortative. Almost right after the administration of the drugs, the functional brain networks started to assume a high variation in the assortativity, being sometimes assortative and other times disassortative, revealing in this frequency band an expressive dynamics of network structural alterations during general anesthesia (see \hyperlink{FIGURE8}{$Figure \cdot 8$, \mbox{Sub-Figure E)}}.

\vspace{-2mm}
 
\subsection{Transitivity}

Blindfolding the monkey has led to an increase in both the variation and the mean values of the transitivity coefficient compared to the time when the monkey had its eyes open (see \hyperlink{FIGURE9}{$Figure \cdot 9$}; \hyperlink{TABLE7}{$Table \cdot 7$}).\\
 
After the anesthetic injection, significant alterations in the transitivity coefficient \citep{newman2001structure} were noticeable on some frequency bands (see \hyperlink{FIGURE9}{$Figure \cdot 9$}).
 
 After administering the anesthetics, the Delta band presented an increase in the transitivity coefficient, with the variation of the values kept almost constant (see \hyperlink{FIGURE9}{$Figure \cdot 9$, \mbox{Sub-Figure A)}.}
 
In the Theta and Alpha bands, during general anesthesia, a decrease in the transitivity and a reduction in the variation of the values occurred (see \hyperlink{FIGURE9}{$Figure \cdot 9$, \mbox{Sub-Figures B and C)}.}
 
In Beta, one minute and a half after the anesthetic injection, an expressive decline occurred. The transitivity coefficient was relatively smaller than when the macaque was awake (eyes open and closed), remaining this way while the monkey was anesthetized (see \hyperlink{FIGURE9}{$Figure \cdot 9$, \mbox{Sub-Figure D)}}.\\

\end{multicols}

\begin{figure}[!h]
\begin{subfigure}{.5\textwidth}
  \centering
  \includegraphics[width=1\linewidth]{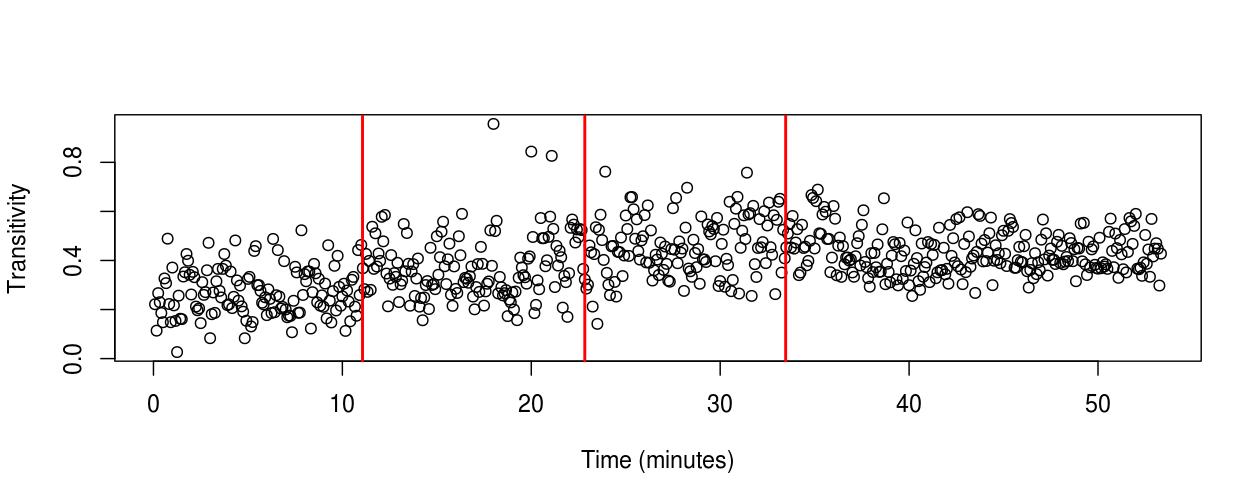}
  \caption{Delta 0-4Hz}
  \label{fig:sfig1}
\end{subfigure}%
\begin{subfigure}{.5\textwidth}
  \centering
  \includegraphics[width=1\linewidth]{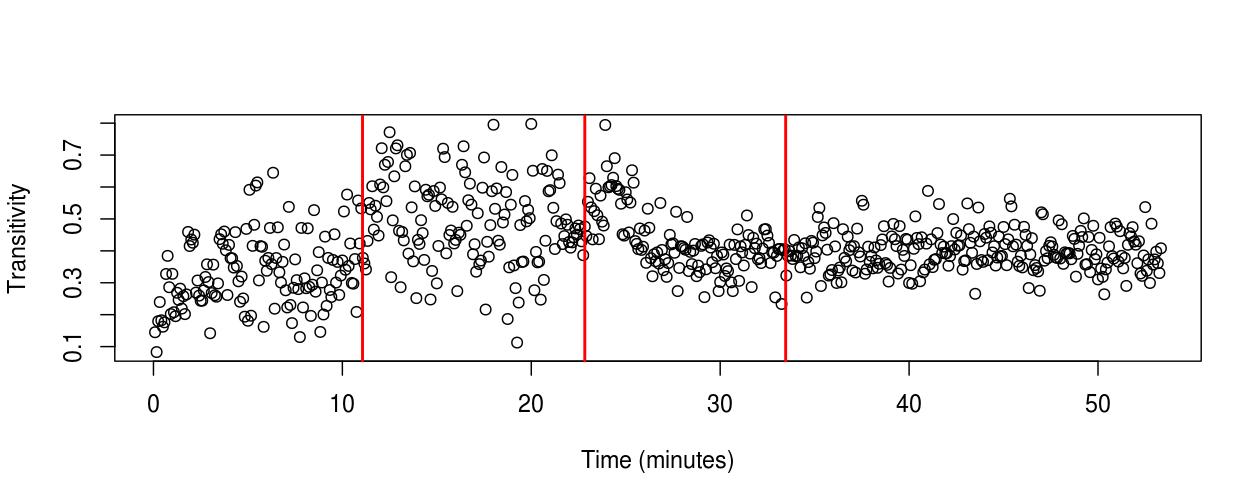}
 \caption{Theta 4-8Hz}
  \label{fig:sfig2}
\end{subfigure}\\
\centering
\begin{subfigure}{.5\textwidth}
\includegraphics[width=1\linewidth]{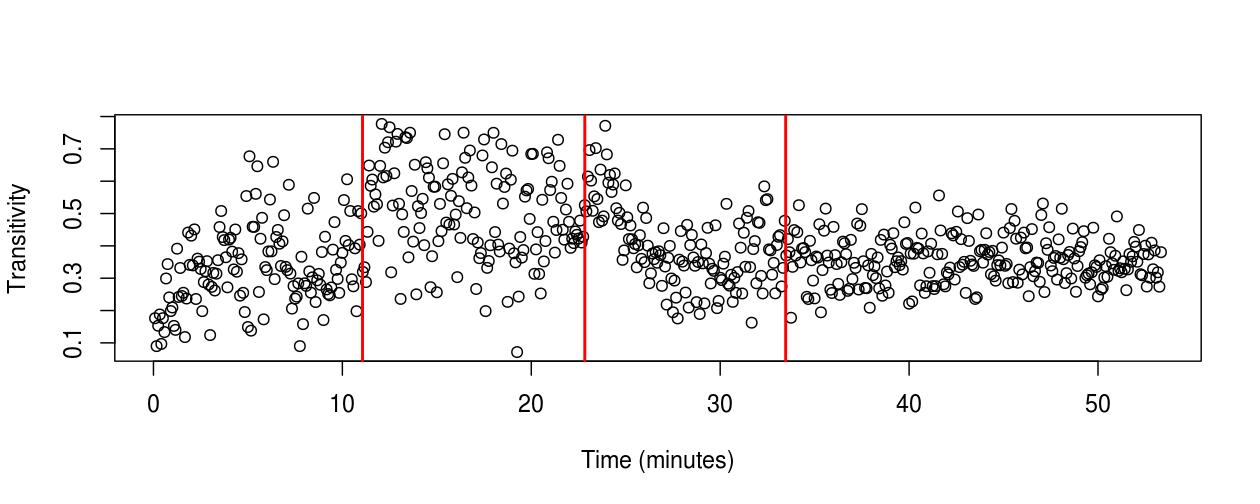}
  \caption{Alpha 8-12Hz}
  \label{fig:sfig3}
\end{subfigure}%
\begin{subfigure}{.5\textwidth}
  \centering
  \includegraphics[width=1\linewidth]{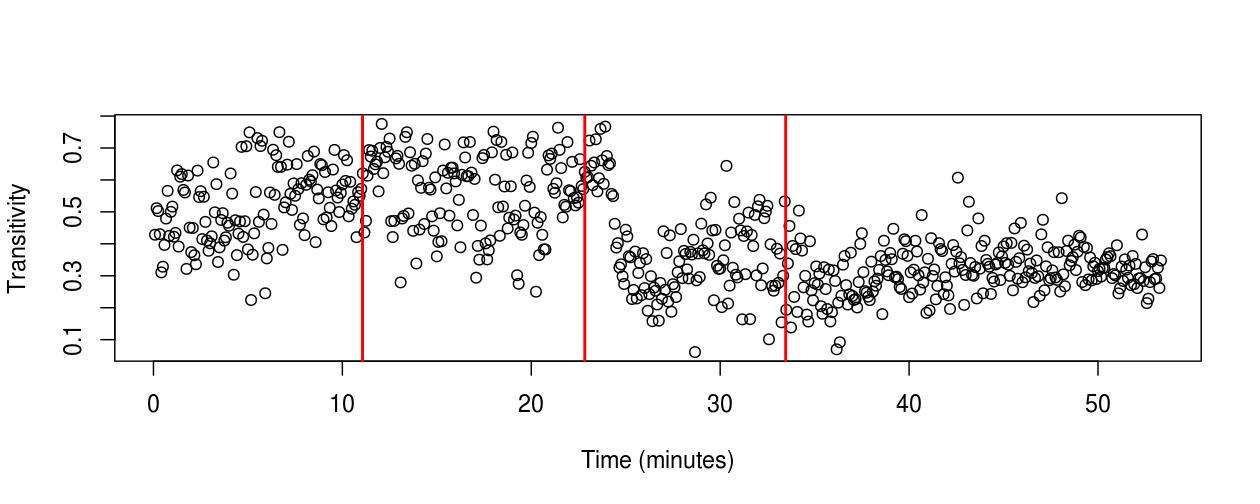}
  \caption{Beta 13-30Hz}
  \label{fig:sfig4}
\end{subfigure}\\
\centering
\begin{subfigure}{.5\textwidth}
\includegraphics[width=1\linewidth]{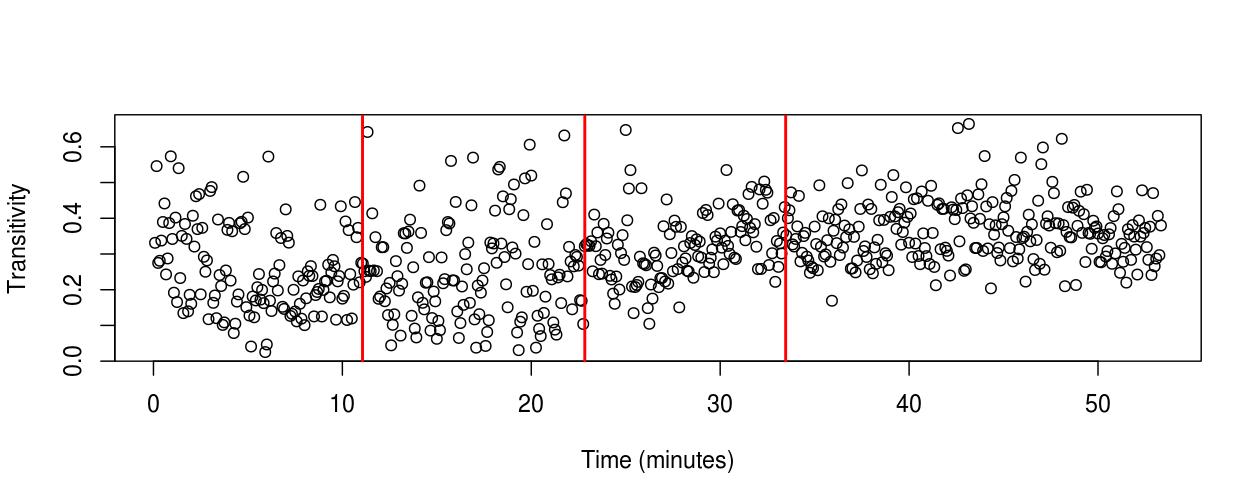}
\caption{Gamma 25-100Hz}
  \label{fig:sfig3}
\end{subfigure}%
\begin{subfigure}{.5\textwidth}
  \centering
  \includegraphics[width=1\linewidth]{white}
  \label{fig:sfig4}
\end{subfigure}
\caption{\textbf{Transitivity}. Vertical axis transitivity coefficient; Horizontal axis time (minutes). At t=11 minutes, the monkey was blindfolded; the first vertical red line represents this event in each sub-figure. Next, at t=23 minutes, the Ketamine-Medetomidine cocktail was injected, represented by the second vertical red line. Finally, the point of loss of consciousness (LOC) was registered at t=33 minutes, indicated by the third vertical red line.}
\label{fig:fig}
\hypertarget{FIGURE9}{}
\end{figure}

\begin{table}[h]
\centering
\caption{Mean, variance (Var), and standard deviation (SD) of the transitivity coefficient on the five physiological frequency bands analyzed and on the three different conditions in which the monkey was exposed during the experiment: awake with eyes open, awake with eyes closed and anesthesia (eyes closed).}
\vspace{0.5cm}
\begin{tabular}{l|lcr|lcr|lcr}
 \hline
\textbf{Transitivity} & \multicolumn{3}{c}{Eyes Open} \vline &\multicolumn{3}{c}{Eyes Closed} \vline &\multicolumn{3}{c}{Anesthesia}\\
\hline
Frequency Band & Mean & Var & SD  & Mean & Var & SD & Mean & Var & SD\\ 
\hline                             
Delta (0-4Hz)   &0,27     &0,01  &0.10   &0,37  &0.02   &0.14     &0.44    &0.01    &0.10   \\
Theta  (4-8Hz)  &0,32     &0,01  &0.10   &0,48  &0.01   &0.10     &0.40    &0.01    &0.10   \\
Alpha (8-12Hz)  &0,32     &0,01  &0.10   &0,49  &0.02   &0,14     &0.35    &0.01    &0.10       \\
Beta (13-30Hz)  &0,51     &0,01  &0.10   &0,55  &0.01   &0,10     &0.32    &0.01    &0.10       \\
Gamma (25-100Hz)&0,26     &0,01  &0.10   &0,25  &0.02   &0,14     &0.36    &0.01    &0.10       
 
\end{tabular}
\hypertarget{TABLE7}{}
\end{table}

\begin{multicols}{2}

In Gamma, a reduction in the values was observed one minute and a half after administering the anesthetics. Later, the transitivity coefficient started to increase. Another remarkable alteration in this frequency band was a smaller variation of the values compared to the period when the monkey was awake \mbox{(see \hyperlink{FIGURE9}{$Figure \cdot 9$, \mbox{Sub-Figure E)}}}.

\enlargethispage{1.5\baselineskip}

\vspace{1.5\baselineskip}

\section{Discussion}

In the experiment, the administration of the Ketamine-Medetomidine cocktail led to alterations in various network properties on the five physiological frequency bands analyzed. The most remarkable alterations occurred in the Theta, Alpha, and Beta bands. Those experimental observations indicate that the Ketamine-Medetomidine cocktail affected more expressively neural activities occurring between 4 and 30Hz. Once the anesthetic agents promoted physiological changes in the animal subject, drastically reducing its cognitive capacity and level of consciousness, suggest that neural processes and activities related to cognition and consciousness may primarily occur within frequencies ranging from 4 to 30 Hz.\\

Changes in several topological properties of the networks occurred within approximately one and a half minutes after administering the anesthetics. This observation of consistent changes occurring rapidly and simultaneously in different properties of the graphs strongly indicates a phase transition in the network's architecture. Furthermore, the fact that phenomena occurred straight after the injection of the anesthetic cocktail provides strong confidence that the changes observed during the experiment are directly related to the physiological effects of the anesthetics on the animal (a rapid and expressive reduction in the level of consciousness).

Those results are exciting according to the perspective of the \textit{Modern Network Science} once they bring empirical evidence involving relations between a change in the behavior of a natural system and alterations in its respective underlying networks. They are in agreement with one of the remarkable premises of complex systems science, which states that the behavior exhibited by a system is intimately related to its structure, that is, to the specific way in which the elements of the system establish interactions.

 
The present research is one of the first papers to report the existence and estimate the structure and dynamics of large-scale functional brain networks respective to induced general anesthesia.
 
\subsection{Alterations on the Network's Properties}

\subsection*{Average degree}

The average degree reflects the average number of connections of the network's vertices \citep{rubinov2010complex}, providing information related to the global connectivity of the graph, indicating how interactive the elements of the system are.
 
 The results obtained in this research revealed that after administering the anesthetics, there was a significant reduction in the average degree respective to the Theta, Alpha, and Beta bands (see \hyperlink{FIGURE1}{$Figure \cdot 1$}; \hyperlink{TABLE1}{$Table \cdot 1$}). From this experimental evidence, it has been verified that the anesthetic induction leads to a significant reduction in cortical connectivity, demonstrating that functional interactions established among various regions of the cortex are impaired during general anesthesia.


\subsection*{ Vertices Degree and Cortical Areas}

Considering the functional brain networks, the degree of each vertex brings information regarding the capacity of the area that it represents to influence or receive influence from other distinct areas \citep{sporns2011networks}. 
 
The results of this experiment demonstrate that in the awake state, high-degree vertices cover all frontal and parietal lobes and temporal areas anterior to the superior temporal sulcus, indicating that in awakening conditions, those cortical regions present high connectivity and functional integration. After administering the anesthetics, highly connected vertices were no longer observed in those cortical areas, evidencing a drastic reduction in functional connectivity. The results reveal that the Ketamine-Medetomidine cocktail mainly affected the prominent connectivity that existed in areas of the secondary associative cortex of the frontal and parietal lobes, strongly suggesting that events of consciousness may be highly dependent on the high functional integration involving these anatomical regions. Such experimental evidence is in accordance with several scientific reports relating the same regions to neural correlates of consciousness. Furthermore, there is experimental evidence that the administration of the general anesthetic Propofol induces a significant decrease in blood flow in areas of the pre-cuneus, cuneus, and posterior cingulate cortex \citep{fiset1999brain}, with the inactivation of the posterior medial part associated with loss of consciousness \citep{kaisti2002effects}. Damasio and Dolan reported that injuries affecting the same cortical areas are related to severe disturbances of cognition and consciousness \citep{damasio1999feeling}. Laureys highlighted that frontoparietal regions are preferentially deactivated in human patients in the vegetative states \citep{laureys2004brain}, with the loss of consciousness of those patients associated with the functional disconnectivity between frontal and parietal regions \citep{laureys1999impaired}.

\subsection*{Average Degree - Cortical Lobes}
The analysis of the average degree of the sub-graphs respective to the frontal, parietal, temporal, and occipital lobes confirms some features observed in the color gradient representing the degree of the vertices over the position of the electrodes. In addition, those results show that on the Theta, Alpha, and Beta\footnote{Theta and Beta frequency bands data not shown.} frequency bands, the anesthetic induction led to an expressive decrease in functional connectivity over the frontal, parietal, and temporal lobes (see \hyperlink{FIGURE3}{$Figure \cdot 3$}; \hyperlink{TABLE2}{$Table \cdot 2$}). 
 
Unlike what happened in the frontal, parietal, and temporal lobes, the average connectivity of the subgraphs respective to the occipital lobe was three times higher\footnote{When compared to the awake state (eyes open).} during general anesthesia (see \hyperlink{TABLE2}{$Table \cdot 2$}). Furthermore, such an event was also observed in the color gradient representing the vertice's degree. These results strongly suggest the presence of coherent neural activity in the occipital lobe during general anesthesia induced by Ketamine and Metomidine.


\subsection*{Average Path Length}

According to Latora and Marchiori, a network's capacity and global efficiency in transmitting information are directly related to the average of its minimum paths, the most efficient networks are those with the shortest paths \citep{latora2001efficient}.

The experimental results demonstrate that the administration of the anesthetics led to a substantial increase in the average path length \citep{costa2007characterization} on the Theta, Alpha, and Beta frequency bands (see \hyperlink{FIGURE4}{$Figure \cdot 4$}; \hyperlink{TABLE3}{$Table \cdot 3$}). Such an increase in the average path length strongly indicates that the overall capacity for transmitting information between multiple cortical areas is significantly reduced during general anesthesia.

\subsection*{Diameter}

The diameter of the network \citep{costa2007characterization}, once related only to the larger geodesic path of the graph, is a less informative measure than the average path length once the latter takes into account all the minimum paths of the network. The diameter reflects the length of the largest minimum path, representing the maximum distance existent on the network\footnote{Assuming that the graph is connected.}.
 
The experimental results reveal a considerable increase in the length of the diameter of the networks on the Theta, Alpha, and Beta bands, which occurred one and a half minutes after the administration of the anesthetics (see \hyperlink{FIGURE5}{$Figure \cdot 5$}). Such a result supports the conclusions obtained by analyzing the average path length, that the global transmission of information is reduced during general anesthesia.

\subsection*{Transitivity}

According to Latora and Marchiori, the local efficiency of information transmission in a network is directly related to its transitivity coefficient. The larger the coefficient, the greater the local efficiency of the network \citep{latora2003economic}.

The results obtained in this study demonstrate that the anesthetics led to a reduction of the transitivity coefficient\footnote{Mainly when compared to the awake state (eyes closed), in Theta, Alpha, and Beta, the most prominent reduction observed in the Beta band.} (see \hyperlink{TABLE7}{$Table \cdot 7$}; \hyperlink{FIGURE9}{$Figure \cdot 9$}). Such a decrease indicates that the transmission of information efficiency is reduced at local levels during the induced state of anesthesia.

\subsection*{Assortativity Coefficient}

The assortativity coefficient \citep{boccaletti2006complex} is related to the preferential attachment between vertices concerning their connectivity degree \citep{boccaletti2006complex}. Alterations in this coefficient reveal structural changes in the graphs, specifically the manner in which the connections between the vertices are established.

The results of this experiment revealed that the anesthetics led to significant alterations in the network's assortativity character. The most accentuated changes were observed in the Delta, Theta, and Gamma bands (see \hyperlink{FIGURE8}{$Figure \cdot 8$}). These results demonstrate structural changes in the neural activity network's organization over those frequency bands.
The graphs corresponding to the Delta and Gamma bands showed no noticeable changes in other network measures after the anesthetic induction process. However, alterations in the assortativity coefficient highlight that the anesthetics led to significant structural rearrangements in the large-scale functional brain networks.

As reported by Costa, the network's assortativity character may influence the dynamic processes supported by the system \citep{costa2007characterization}. The fact that the networks respective to the induced state of anesthesia assumed a predominantly assortative character might imply a higher instability of those networks \citep{brede2005assortative} and a reduction in their synchronization capacity \citep{di2005synchronization}.
 
\subsection*{Intermediation Degree and Cortical Areas}
\vspace{+1.0\baselineskip}
 
The betweenness centrality degree of a vertex \citep{freeman1979centrality} is related to the relevance of this vertex, given its participation in the minimum paths of the network. The larger the number of minimum paths passing through a node, the more significant the impact on the network integration performed by this vertex.
The analysis of the relationship between the intermediation degree of the vertices and the position of the electrodes over the cortex provides an estimate of how information transmission is performed on the network, highlighting the areas that monopolize the flux of information and play an essential role in functional integration.\\
 
The results obtained experimentally demonstrate that during awake conditions (eyes open and closed), the vast majority of the vertices of the graph had a high intermediation degree (see \hyperlink{FIGURE7}{$Figure \cdot 7$}), indicating that the flow of information on the network is in a certain way distributed between multiple cortical areas. Those results suggest that during awake conditions, the vast majority of the cortex gives support and is possibly actively involved in information transmission. Another remarkable characteristic is that high intermediation vertices extended, covering continuously large cortical areas, without being separated by regions characterized by vertices possessing a low intermediation degree (see \hyperlink{FIGURE7}{$Figure \cdot 7$, \mbox{Sub-Figures A - E)}}.

After administering the anesthetics, a significant change concerning the intermediation degree of the vertices and their respective cortical areas occurred (see \hyperlink{FIGURE7}{$Figure \cdot 7$}). It was possible to observe that some areas monopolized the integration of the network and that the vast majority of the cortex presented a reduced degree of intermediation, revealing that the structure of the network did not provide the distribution of the flow of information over ``the entire'' cortex anymore.
Another remarkable characteristic observed was the discontinuity of the occupancy of the areas having a high degree of intermediation, being those physically segregated by extensive regions marked by vertices possessing a low intermediation degree (see \hyperlink{FIGURE7}{$Figure \cdot 7$, \mbox{Sub-Figures P - T)}}. From the analysis of the results, it is possible to infer the hypothesis that activity and neural processes associated with conscious experiences may require the involvement of a significant portion of the cortex in the integration of the entire network.
 
\subsection*{Average Betweenness Centrality Degree}

A direct association between the average intermediation degree of the vertices and functional or structural properties of the network is not trivial once distinct alterations in the graph's topology may lead to similar changes in the mean intermediation degree of the vertices. Besides not being able to comprehend straightforwardly the meanings and implications of the alterations observed on this property, the existence of changes confirms that structural alterations have occurred on the graphs.
 
In this experiment, after administering the anesthetics, noticeable changes were observed in the average betweenness centrality degree in the Theta, Alpha, Beta, and Gamma bands (see \hyperlink{TABLE5}{$Table \cdot 5$}; \hyperlink{FIGURE6}{$Figure \cdot 6$}), confirming the occurrence of structural rearrangements in the large-scale functional brain networks during general anesthesia.

\subsection{ Alterations  due to Anesthesia on Small-World  Architecture}

Regarding the five frequency bands analyzed, the most significant alterations observed during the anesthetic induction occurred on the networks respective to the Theta, Alpha, and Beta bands. In general, those frequency bands presented mainly the same kind of alterations:

\begin{itemize}

\item \textbf{An expressive reduction on the average degree of the vertices.}

\item \textbf{A considerable increase in the average path length.
}
\item \textbf{A decrease in the transitivity coefficient.}
 
\end{itemize}
The considerable reduction in the average degree and transitivity and the significant increase in the average path length have direct consequences, impacting the small world architecture \citep{watts1998collective} observed during the awake state. According to Olaf Sporns, the small-world architecture, characterized by high values of the transitivity coefficient and a small average path length, provides a structural substrate of great relevance, important in many aspects of the brain's functional organization. This architecture supports processing information segregated locally and integrated globally \citep{sporns2011networks}. The small world architecture is also considered to promote the efficiency of transmission and processing of information, the wiring economy, and the support of diverse and complex network dynamics \citep{bullmore2012economy}.

Several authors consider segregation and integration as two of the main principles of the organization of activities that occur in the cerebral cortex \citep{zeki1978functional,zeki1988functional,
tononi1994measure,tononi1998consciousness,friston2002beyond,
friston2005models,friston2009modalities}. Sporns highlights that the balance between those two properties constitutes a key mechanism necessary for the brain to perform its activities \citep{sporns2011networks}.
 
 The results of this study demonstrate that the anesthetic induction had most significantly impacted two network properties over the Theta, Alpha, and Beta frequency bands. These properties were the integration capacity at global levels (an increase in the average path length) and the integration capacity at local levels (a decrease in the transitivity coefficient). Those factors have a direct consequence on the breakdown of the small-world architecture that was observed in the awake state. These experimental results support hypotheses pointed out by many authors, relating the loss of fundamental structural properties of functional networks to alterations and suppression of brain functions.

\subsection*{Alterations Observed After the Placement of the Blindfold}

Several alterations in the topology and dynamic behavior of the properties of the large-scale functional brain networks were observed after placing a blindfold over the eyes of the macaque. It is possible from the figures and graphs to note clear distinctions regarding the time when the monkey had its eyes open and when it had its eyes closed.\\
 
\noindent \ The researchers who conducted the experiments and provided the database of the neural activity records did not provide enough information regarding the placement of the blindfold on the state of the animal. The effects may depend on several factors, such as the macaque's conditioning to the experimental settings, the animal's familiarity with the researchers, the handling techniques used during the experiment, and other variables. For example, placing a blindfold over the eyes of the monkey and restraining the animal in a chair might have led to a calm and relaxed state or triggered fear and apprehension. Thus, conclusions regarding the relationship between the network measures and the specific state in which the monkey was can only be drawn under an experiment's detailed description.
The fact that alterations were observed in the network measures immediately after blindfolding the monkey demonstrates that the functional networks are dependent on the conditions presented to the animal. Those observed phenomena follow some of the suppositions from Olaf Sporns, who stated that functional connectivity might vary considerably over time, modulated by demands due to different types of activities or due to different sensory stimuli presented \citep{sporns2011networks}.

\subsection*{Point of Loss of Consciousness (LOC)}

The scientists who conducted the experimental procedures and recorded the data reported the occurrence of the loss of consciousness point (LOC) at the moment when the animal ceased to respond to stimuli, such as touching the nostrils and hands. However, no prominent alterations in network properties were observed at the (LOC),  registered by the researchers about ten minutes after administering the anesthetics. Therefore, the awake state probably ended when several changes in the properties of the networks were observed, which occurred less than two minutes after the anesthetic injection. However, after that time, the monkey was still able to exhibit involuntary reactions to stimuli, being in an intermediary state between deep sedation and general anesthesia, probably in similar conditions as the pharmacological effects of the administration of Ketamine in human patients are described \citep{bergman1999ketamine}.
 
 \section{Conclusions}

The supposed view of general anesthesia being given by a "whole brain shutdown" is not supported by the experimental findings of this study. Instead, it was possible to observe that during the Ketamine-medetomidine-induced anesthesia, the brain entered into a highly specific, complex, and dynamic state.

By modeling the interactions established among several cortical areas through graphs and complex networks, it was possible to verify that the networks corresponding to the induced state of anesthesia were structurally distinct from the networks respective to the awake state. Those results reveal that anesthetics can impact several properties of the large-scale functional brain networks, resulting in graphs of different architectures. Furthermore, the changes observed in the experiment indicate that the behavior exhibited and the processes supported by the brain may be directly related to the structural properties of large-scale functional networks.
It was possible to conclude that functional neural activities are dynamic, as significant changes were observed in the network measures over short time intervals. Observing the degree of the vertices through a color gradient also reveals consistent pattern changes occurring within a few seconds. Such evidence demonstrates that functional neural activities are not static; they reveal that the brain's functional activity organization is remarkably dynamic\footnote{In both awake and anesthetized conditions.}.
 
The Ketamine-Medetomidine cocktail did not alter the functional activities in only some specific and restricted areas of the cortex; a general change in the state of the brain in which all the cortex presented alterations in functional connectivity occurred.

The most remarkable changes observed in the large-scale functional brain networks due to the anesthetic induction were an accentuated decrease in functional connectivity and a reduction in the capacities of integration of the networks at both local and global levels.

From the characterization of the functional networks during the anesthetic induction process, we verified a transition between the awake and anesthetized states. The transition occurred in a pretty fast manner, taking about 20 to 30 seconds.

\subsubsection*{Financial Support}

This research was partially financed by CAPES (Coordena\c{c}\~ao de Aperfei\c{c}oamento de Pessoal de N\'ivel Superior).

During part of the development of the study, the author used the structure of the Laboratory Vision-eScience at Universidade de S\~ao Paulo, a laboratory supported by FAPESP, CNPq, CAPES, NAP-eScience, PRP, and USP.

\bibliography{dissertacao}

\begin{thebibliography}{60}
\providecommand{\natexlab}[1]{#1}
\providecommand{\url}[1]{\texttt{#1}}
\expandafter\ifx\csname urlstyle\endcsname\relax
  \providecommand{\doi}[1]{doi: #1}\else
  \providecommand{\doi}{doi: \begingroup \urlstyle{rm}\Url}\fi

\bibitem[Alkire(1998)]{alkire1998quantitative}
Michael~T Alkire.
\newblock Quantitative eeg correlations with brain glucose metabolic rate
  during anesthesia in volunteers.
\newblock \emph{Anesthesiology}, 89\penalty0 (2):\penalty0 323--333, 1998.

\bibitem[Alkire et~al.(1995)Alkire, Haier, Barker, Shah, Wu, and
  Kao]{alkire1995cerebral}
Michael~T Alkire, Richard~J Haier, Steven~J Barker, Nitin~K Shah, Joseph~C Wu,
  and Y~James Kao.
\newblock Cerebral metabolism during propofol anesthesia in humans studied with
  positron emission tomography.
\newblock \emph{Anesthesiology}, 82\penalty0 (2):\penalty0 393--403, 1995.

\bibitem[Alkire et~al.(2000)Alkire, Haier, and Fallon]{alkire2000toward}
MT~Alkire, RJ~Haier, and JH~Fallon.
\newblock Toward a unified theory of narcosis: brain imaging evidence for a
  thalamocortical switch as the neurophysiologic basis of anesthetic-induced
  unconsciousness.
\newblock \emph{Consciousness and cognition}, 9\penalty0 (3):\penalty0
  370--386, 2000.

\bibitem[Bergman(1999)]{bergman1999ketamine}
Stewart~A Bergman.
\newblock Ketamine: review of its pharmacology and its use in pediatric
  anesthesia.
\newblock \emph{Anesthesia progress}, 46\penalty0 (1):\penalty0 10, 1999.

\bibitem[Boccaletti et~al.(2006)Boccaletti, Latora, Moreno, Chavez, and
  Hwang]{boccaletti2006complex}
Stefano Boccaletti, Vito Latora, Yamir Moreno, Martin Chavez, and D-U Hwang.
\newblock Complex networks: Structure and dynamics.
\newblock \emph{Physics reports}, 424\penalty0 (4):\penalty0 175--308, 2006.

\bibitem[Brede and Sinha(2005)]{brede2005assortative}
Markus Brede and Sitabhra Sinha.
\newblock Assortative mixing by degree makes a network more unstable.
\newblock \emph{arXiv preprint cond-mat/0507710}, 2005.

\bibitem[Bullmore and Sporns(2009)]{bullmore2009complex}
Ed~Bullmore and Olaf Sporns.
\newblock Complex brain networks: graph theoretical analysis of structural and
  functional systems.
\newblock \emph{Nature Reviews Neuroscience}, 10\penalty0 (3):\penalty0
  186--198, 2009.

\bibitem[Bullmore and Sporns(2012)]{bullmore2012economy}
Ed~Bullmore and Olaf Sporns.
\newblock The economy of brain network organization.
\newblock \emph{Nature Reviews Neuroscience}, 13\penalty0 (5):\penalty0
  336--349, 2012.

\bibitem[Costa et~al.(2007)Costa, Rodrigues, Travieso, and
  Villas~Boas]{costa2007characterization}
L~da~F Costa, Francisco~A Rodrigues, Gonzalo Travieso, and Paulino~Ribeiro
  Villas~Boas.
\newblock Characterization of complex networks: A survey of measurements.
\newblock \emph{Advances in Physics}, 56\penalty0 (1):\penalty0 167--242, 2007.

\bibitem[Cui et~al.(2008)Cui, Xu, Bressler, Ding, and Liang]{cui2008bsmart}
Jie Cui, Lei Xu, Steven~L Bressler, Mingzhou Ding, and Hualou Liang.
\newblock Bsmart: a matlab/c toolbox for analysis of multichannel neural time
  series.
\newblock \emph{Neural Networks}, 21\penalty0 (8):\penalty0 1094--1104, 2008.

\bibitem[Damasio and Dolan(1999)]{damasio1999feeling}
Antonio Damasio and Raymond~J Dolan.
\newblock The feeling of what happens.
\newblock \emph{Nature}, 401\penalty0 (6756):\penalty0 847--847, 1999.

\bibitem[di~Bernardo et~al.(2005)di~Bernardo, Garofalo, and
  Sorrentino]{di2005synchronization}
Mario di~Bernardo, Franco Garofalo, and Francesco Sorrentino.
\newblock Synchronization of degree correlated physical networks.
\newblock \emph{arXiv preprint cond-mat/0506236}, 2005.

\bibitem[Fiset et~al.(1999)Fiset, Paus, Daloze, Plourde, Meuret, Bonhomme,
  Hajj-Ali, Backman, and Evans]{fiset1999brain}
Pierre Fiset, Tom{\'a}s Paus, Thierry Daloze, Gilles Plourde, Pascal Meuret,
  Vincent Bonhomme, Nadine Hajj-Ali, Steven~B Backman, and Alan~C Evans.
\newblock Brain mechanisms of propofol-induced loss of consciousness in humans:
  a positron emission tomographic study.
\newblock \emph{The Journal of neuroscience}, 19\penalty0 (13):\penalty0
  5506--5513, 1999.

\bibitem[Flohr(1995)]{flohr1995information}
Hans Flohr.
\newblock An information processing theory of anaesthesia.
\newblock \emph{Neuropsychologia}, 33\penalty0 (9):\penalty0 1169--1180, 1995.

\bibitem[Freeman(1979)]{freeman1979centrality}
Linton~C Freeman.
\newblock Centrality in social networks conceptual clarification.
\newblock \emph{Social networks}, 1\penalty0 (3):\penalty0 215--239, 1979.

\bibitem[Friston(2002)]{friston2002beyond}
Karl Friston.
\newblock Beyond phrenology: what can neuroimaging tell us about distributed
  circuitry?
\newblock \emph{Annual review of neuroscience}, 25\penalty0 (1):\penalty0
  221--250, 2002.

\bibitem[Friston(1994)]{friston1994functional}
Karl~J Friston.
\newblock Functional and effective connectivity in neuroimaging: a synthesis.
\newblock \emph{Human brain mapping}, 2\penalty0 (1-2):\penalty0 56--78, 1994.

\bibitem[Friston(2005)]{friston2005models}
Karl~J Friston.
\newblock Models of brain function in neuroimaging.
\newblock \emph{Annu. Rev. Psychol.}, 56:\penalty0 57--87, 2005.

\bibitem[Friston(2009)]{friston2009modalities}
Karl~J Friston.
\newblock Modalities, modes, and models in functional neuroimaging.
\newblock \emph{Science}, 326\penalty0 (5951):\penalty0 399--403, 2009.

\bibitem[Friston et~al.(1993)Friston, Frith, Liddle, and
  Frackowiak]{friston1993functional}
KJ~Friston, CD~Frith, PF~Liddle, and RSJ Frackowiak.
\newblock Functional connectivity: the principal-component analysis of large
  (pet) data sets.
\newblock \emph{Journal of Cerebral Blood Flow \& Metabolism}, 13\penalty0
  (1):\penalty0 5--14, 1993.

\bibitem[Fukushima et~al.(2014)Fukushima, Saunders, Mullarkey, Doyle, Mishkin,
  and Fujii]{fukushima2014electrocorticographic}
Makoto Fukushima, Richard~C Saunders, Matthew Mullarkey, Alexandra~M Doyle,
  Mortimer Mishkin, and Naotaka Fujii.
\newblock An electrocorticographic electrode array for simultaneous recording
  from medial, lateral, and intrasulcal surface of the cortex in macaque
  monkeys.
\newblock \emph{Journal of neuroscience methods}, 233:\penalty0 155--165, 2014.

\bibitem[Granger(1969)]{granger1969investigating}
Clive~WJ Granger.
\newblock Investigating causal relations by econometric models and
  cross-spectral methods.
\newblock \emph{Econometrica: Journal of the Econometric Society}, pages
  424--438, 1969.

\bibitem[Green et~al.(2011)Green, Roback, Kennedy, and
  Krauss]{green2011clinical}
Steven~M Green, Mark~G Roback, Robert~M Kennedy, and Baruch Krauss.
\newblock Clinical practice guideline for emergency department ketamine
  dissociative sedation: 2011 update.
\newblock \emph{Annals of emergency medicine}, 57\penalty0 (5):\penalty0
  449--461, 2011.

\bibitem[Hameroff(2006)]{hameroff2006entwined}
Stuart Hameroff.
\newblock The entwined mysteries of anesthesia and consciousness.
\newblock \emph{Anesthesiology}, 105\penalty0 (2):\penalty0 400--12, 2006.

\bibitem[Hameroff et~al.(1998)Hameroff, Kaszniak, and
  Scott]{hameroff1998toward}
Stuart~R Hameroff, Alfred~W Kaszniak, and Alwyn Scott.
\newblock \emph{Toward a science of consciousness II: The second Tucson
  discussions and debates}, volume~2.
\newblock Mit Press, 1998.

\bibitem[Hamilton(1989)]{hamilton1989new}
James~D Hamilton.
\newblock A new approach to the economic analysis of nonstationary time series
  and the business cycle.
\newblock \emph{Econometrica: Journal of the Econometric Society}, pages
  357--384, 1989.

\bibitem[Imas et~al.(2005)Imas, Ropella, Ward, Wood, and
  Hudetz]{imas2005volatile}
Olga~A Imas, Kristina~M Ropella, B~Douglas Ward, James~D Wood, and Anthony~G
  Hudetz.
\newblock Volatile anesthetics disrupt frontal-posterior recurrent information
  transfer at gamma frequencies in rat.
\newblock \emph{Neuroscience letters}, 387\penalty0 (3):\penalty0 145--150,
  2005.

\bibitem[Iriki and Sakura(2008)]{iriki2008neuroscience}
Atsushi Iriki and Osamu Sakura.
\newblock The neuroscience of primate intellectual evolution: natural selection
  and passive and intentional niche construction.
\newblock \emph{Philosophical Transactions of the Royal Society B: Biological
  Sciences}, 363\penalty0 (1500):\penalty0 2229--2241, 2008.

\bibitem[John and Prichep(2005)]{john2005anesthetic}
E~Roy John and Leslie~S Prichep.
\newblock The anesthetic cascade: a theory of how anesthesia suppresses
  consciousness.
\newblock \emph{Anesthesiology}, 102\penalty0 (2):\penalty0 447, 2005.

\bibitem[Kaisti et~al.(2002)Kaisti, Mets{\"a}honkala, Ter{\"a}s, Oikonen,
  Aalto, J{\"a}{\"a}skel{\"a}inen, Hinkka, and Scheinin]{kaisti2002effects}
Kaike~K Kaisti, Liisa Mets{\"a}honkala, Mika Ter{\"a}s, Vesa Oikonen, Sargo
  Aalto, Satu J{\"a}{\"a}skel{\"a}inen, Susanna Hinkka, and Harry Scheinin.
\newblock Effects of surgical levels of propofol and sevoflurane anesthesia on
  cerebral blood flow in healthy subjects studied with positron emission
  tomography.
\newblock \emph{Anesthesiology}, 96\penalty0 (6):\penalty0 1358--1370, 2002.

\bibitem[Kennedy and Norman(2005)]{kennedy2005don}
Donald Kennedy and Colin Norman.
\newblock What don't we know?
\newblock \emph{Science}, 309\penalty0 (5731):\penalty0 75--75, 2005.

\bibitem[Kwiatkowski et~al.(1992)Kwiatkowski, Phillips, Schmidt, and
  Shin]{kwiatkowski1992testing}
Denis Kwiatkowski, Peter~CB Phillips, Peter Schmidt, and Yongcheol Shin.
\newblock Testing the null hypothesis of stationarity against the alternative
  of a unit root: How sure are we that economic time series have a unit root?
\newblock \emph{Journal of econometrics}, 54\penalty0 (1):\penalty0 159--178,
  1992.

\bibitem[Latora and Marchiori(2001)]{latora2001efficient}
Vito Latora and Massimo Marchiori.
\newblock Efficient behavior of small-world networks.
\newblock \emph{Physical review letters}, 87\penalty0 (19):\penalty0 198701,
  2001.

\bibitem[Latora and Marchiori(2003)]{latora2003economic}
Vito Latora and Massimo Marchiori.
\newblock Economic small-world behavior in weighted networks.
\newblock \emph{The European Physical Journal B-Condensed Matter and Complex
  Systems}, 32\penalty0 (2):\penalty0 249--263, 2003.

\bibitem[Laureys et~al.(1999)Laureys, Goldman, Phillips, Van~Bogaert, Aerts,
  Luxen, Franck, and Maquet]{laureys1999impaired}
Steven Laureys, Serge Goldman, Christophe Phillips, Patrick Van~Bogaert,
  Jo{\"e}l Aerts, Andr{\'e} Luxen, Georges Franck, and Pierre Maquet.
\newblock Impaired effective cortical connectivity in vegetative state:
  preliminary investigation using pet.
\newblock \emph{Neuroimage}, 9\penalty0 (4):\penalty0 377--382, 1999.

\bibitem[Laureys et~al.(2004)Laureys, Owen, and Schiff]{laureys2004brain}
Steven Laureys, Adrian~M Owen, and Nicholas~D Schiff.
\newblock Brain function in coma, vegetative state, and related disorders.
\newblock \emph{The Lancet Neurology}, 3\penalty0 (9):\penalty0 537--546, 2004.

\bibitem[Lewis et~al.(2012)Lewis, Weiner, Mukamel, Donoghue, Eskandar, Madsen,
  Anderson, Hochberg, Cash, Brown, et~al.]{lewis2012rapid}
Laura~D Lewis, Veronica~S Weiner, Eran~A Mukamel, Jacob~A Donoghue, Emad~N
  Eskandar, Joseph~R Madsen, William~S Anderson, Leigh~R Hochberg, Sydney~S
  Cash, Emery~N Brown, et~al.
\newblock Rapid fragmentation of neuronal networks at the onset of
  propofol-induced unconsciousness.
\newblock \emph{Proceedings of the National Academy of Sciences}, 109\penalty0
  (49):\penalty0 E3377--E3386, 2012.

\bibitem[Mashour(2004)]{mashour2004consciousness}
George~A Mashour.
\newblock Consciousness unbound: toward a paradigm of general anesthesia.
\newblock 2004.

\bibitem[Mashour(2006)]{mashour2006integrating}
George~A Mashour.
\newblock Integrating the science of consciousness and anesthesia.
\newblock \emph{Anesthesia \& Analgesia}, 103\penalty0 (4):\penalty0 975--982,
  2006.

\bibitem[Mitchell(2009)]{mitchell2009complexity}
Melanie Mitchell.
\newblock \emph{Complexity: A guided tour}.
\newblock Oxford University Press, 2009.

\bibitem[Nagasaka et~al.(2011)Nagasaka, Shimoda, and
  Fujii]{nagasaka2011multidimensional}
Yasuo Nagasaka, Kentaro Shimoda, and Naotaka Fujii.
\newblock Multidimensional recording (mdr) and data sharing: an ecological open
  research and educational platform for neuroscience.
\newblock \emph{PloS one}, 6\penalty0 (7):\penalty0 e22561, 2011.

\bibitem[Newman(2001)]{newman2001structure}
Mark~EJ Newman.
\newblock The structure of scientific collaboration networks.
\newblock \emph{Proceedings of the National Academy of Sciences}, 98\penalty0
  (2):\penalty0 404--409, 2001.

\bibitem[Newman(2003)]{newman2003structure}
Mark~EJ Newman.
\newblock The structure and function of complex networks.
\newblock \emph{SIAM review}, 45\penalty0 (2):\penalty0 167--256, 2003.

\bibitem[Rubinov and Sporns(2010)]{rubinov2010complex}
Mikail Rubinov and Olaf Sporns.
\newblock Complex network measures of brain connectivity: uses and
  interpretations.
\newblock \emph{Neuroimage}, 52\penalty0 (3):\penalty0 1059--1069, 2010.

\bibitem[Schwartz et~al.(2010)Schwartz, Brown, Lydic, and
  Schiff]{schwartz2010general}
Robert~S Schwartz, Emery~N Brown, Ralph Lydic, and Nicholas~D Schiff.
\newblock General anesthesia, sleep, and coma.
\newblock \emph{New England Journal of Medicine}, 363\penalty0 (27):\penalty0
  2638--2650, 2010.

\bibitem[Seth(2010)]{seth2010matlab}
Anil~K Seth.
\newblock A matlab toolbox for granger causal connectivity analysis.
\newblock \emph{Journal of neuroscience methods}, 186\penalty0 (2):\penalty0
  262--273, 2010.

\bibitem[Seth and Edelman(2007)]{seth2007distinguishing}
Anil~K Seth and Gerald~M Edelman.
\newblock Distinguishing causal interactions in neural populations.
\newblock \emph{Neural computation}, 19\penalty0 (4):\penalty0 910--933, 2007.

\bibitem[Sporns(2011)]{sporns2011networks}
Olaf Sporns.
\newblock \emph{Networks of the Brain}.
\newblock MIT press, 2011.

\bibitem[Stam and Van~Straaten(2012)]{stam2012organization}
CJ~Stam and ECW Van~Straaten.
\newblock The organization of physiological brain networks.
\newblock \emph{Clinical Neurophysiology}, 123\penalty0 (6):\penalty0
  1067--1087, 2012.

\bibitem[Stam and Reijneveld(2007)]{stam2007graph}
Cornelis~J Stam and Jaap~C Reijneveld.
\newblock Graph theoretical analysis of complex networks in the brain.
\newblock \emph{Nonlinear biomedical physics}, 1\penalty0 (1):\penalty0 3,
  2007.

\bibitem[Strogatz(2001)]{strogatz2001exploring}
Steven~H Strogatz.
\newblock Exploring complex networks.
\newblock \emph{Nature}, 410\penalty0 (6825):\penalty0 268--276, 2001.

\bibitem[Tononi and Edelman(1998)]{tononi1998consciousness}
Giulio Tononi and Gerald~M Edelman.
\newblock Consciousness and complexity.
\newblock \emph{Science}, 282\penalty0 (5395):\penalty0 1846--1851, 1998.

\bibitem[Tononi et~al.(1994)Tononi, Sporns, and Edelman]{tononi1994measure}
Giulio Tononi, Olaf Sporns, and Gerald~M Edelman.
\newblock A measure for brain complexity: relating functional segregation and
  integration in the nervous system.
\newblock \emph{Proceedings of the National Academy of Sciences}, 91\penalty0
  (11):\penalty0 5033--5037, 1994.

\bibitem[Uhrig et~al.(2014)Uhrig, Dehaene, and Jarraya]{uhrig2014cerebral}
L~Uhrig, S~Dehaene, and B~Jarraya.
\newblock Cerebral mechanisms of general anesthesia.
\newblock In \emph{Annales francaises d'anesthesie et de reanimation},
  volume~33, pages 72--82. Elsevier, 2014.

\bibitem[Watts and Strogatz(1998)]{watts1998collective}
Duncan~J Watts and Steven~H Strogatz.
\newblock Collective dynamics of ''small-world'' networks.
\newblock \emph{Nature}, 393\penalty0 (6684):\penalty0 440--442, 1998.

\bibitem[White and Alkire(2003)]{white2003impaired}
Nathan~S White and Michael~T Alkire.
\newblock Impaired thalamocortical connectivity in humans during
  general-anesthetic-induced unconsciousness.
\newblock \emph{Neuroimage}, 19\penalty0 (2):\penalty0 402--411, 2003.

\bibitem[Yanagawa et~al.(2013)Yanagawa, Chao, Hasegawa, and
  Fujii]{yanagawa2013large}
Toru Yanagawa, Zenas~C Chao, Naomi Hasegawa, and Naotaka Fujii.
\newblock Large-scale information flow in conscious and unconscious states: an
  ecog study in monkeys.
\newblock \emph{PloS one}, 8\penalty0 (11):\penalty0 e80845, 2013.

\bibitem[Young et~al.(1999)Young, Schilling, Skeans, and
  Ritacco]{young1999short}
SS~Young, AM~Schilling, S~Skeans, and G~Ritacco.
\newblock Short duration anaesthesia with medetomidine and ketamine in
  cynomolgus monkeys.
\newblock \emph{Laboratory animals}, 33\penalty0 (2):\penalty0 162--168, 1999.

\bibitem[Zeki and Shipp(1988)]{zeki1988functional}
Semir Zeki and Stewart Shipp.
\newblock The functional logic of cortical connections.
\newblock \emph{Nature}, 1988.

\bibitem[Zeki(1978)]{zeki1978functional}
Semir~M Zeki.
\newblock Functional specialisation in the visual cortex of the rhesus monkey.
\newblock \emph{Nature}, 274\penalty0 (5670):\penalty0 423--428, 1978.

\end{thebibliography}

\end{multicols}

\end{document}